\input harvmac
\input epsf

%%%%%%%%%%%%%  DEFINITIONS %%%%%%%%%%%%%%%%%%%%%%%%%%%%%%%%%%%

\def\figin{\epsfcheck\figin}\def\figins{\epsfcheck\figins}
\def\epsfcheck{\ifx\epsfbox\UnDeFiNeD
\message{(NO epsf.tex, FIGURES WILL BE IGNORED)}
\gdef\figin##1{\vskip2in}\gdef\figins##1{\hskip.5in}% blank space instead
\else\message{(FIGURES WILL BE INCLUDED)}%
\gdef\figin##1{##1}\gdef\figins##1{##1}\fi}
\def\DefWarn#1{}
\def\figinsert{\goodbreak\topinsert}
\def\ifig#1#2#3#4{\DefWarn#1\xdef#1{fig.~\the\figno}
\writedef{#1\leftbracket fig.\noexpand~\the\figno}%
\figinsert\figin{\centerline{\epsfxsize=#3mm \epsfbox{#2}}}
\bigskip\medskip\centerline{\vbox{\baselineskip12pt
\advance\hsize by -1truein\noindent\footnotefont{\sl Fig.~\the\figno:}\sl\ #4}}
\bigskip\endinsert\noindent\global\advance\figno by1}

\def\l{\lambda}
\def\L{\Lambda}

\def\F{\Phi}

\def\inv{^{-1}}
\def\Tr{{\rm Tr}}
\def\hf{{1\over 2}}

\def\cF{{\cal F}}

\def\cW{{\cal W}}

\def\({\bigl(}
\def\){\bigr)}
\def\<{\langle\,}
\def\>{\,\rangle}
\def\]{\right]}
\def\[{\left[}
%%%%%%%%%%%%%%%%%%  Volodya's definitions   %%%%%%%%%%%%%%%

\def\CN {{\cal N }}

\def\p {\partial}

\def\inbar{\,\vrule height1.5ex width.4pt depth0pt}
\def\IB{\relax{\rm I\kern-.18em B}}
\def\IC{\relax\hbox{$\inbar\kern-.3em{\rm C}$}}
\def\IP{\relax{\rm I\kern-.18em P}}
\def\IR{\relax{\rm I\kern-.18em R}}

\def\hf{{1\over 2}}

\def\F{\Phi}

\def\p{\partial}

% central charge

\def\tr{\mathop{\rm tr}\nolimits}
\def\Tr{\mathop{\rm tr}\nolimits}

\def\Dsl{\,\raise.15ex\hbox{/}\mkern-13mu D}
\def\Hsl{\,\raise.15ex\hbox{/}\mkern-12.5mu H}
\def\Lsl{\,\raise.15ex\hbox{/}\mkern-13mu L}
\def\hsl{\raise.15ex\hbox{/}\kern-.57em h}
\def\omegasl{\raise.15ex\hbox{/}\kern-.57em\omega}

\def\BK{{\bf K}}
\def\BE{{\bf E}}

\def\bar{\overline}
\def\BC{{\bf C}}

%%%%%%%%%%%%%%%%%%%%  REFERENCES  %%%%%%%%%%%%%%%%%%%%%%%%%%

\lref\CIV{F.~Cachazo, K.~A.~Intriligator and C.~Vafa,
``A large N duality via a geometric transition,''
Nucl.\ Phys.\ B {\bf 603}, 3 (2001), hep-th/0103067.}
\lref\DVone{R.~ Dijkgraaf, C.~ Vafa,
``Matrix Models, Topological Strings, and Supersymmetric Gauge Theories,''
hep-th/0206255.}
\lref\DVtwo{R.~ Dijkgraaf, C.~ Vafa,
``On Geometry and Matrix Models,'' hep-th/0207106.}
\lref\DVthree{R.~ Dijkgraaf, C.~ Vafa, ``A Perturbative Window
into Non-Perturbative Physics,'' hep-th/0208048.}
\lref\KKN{V.A.~ Kazakov, I.K.~ Kostov, N.~Nekrasov,
``D-particles, Matrix Integrals and KP hierachy,''
Nucl.Phys. {\bf B557} (1999) 413-442.}
\lref\HKK{J. Hoppe, V. A. Kazakov, I. K. Kostov,
``Dimensionally Reduced SYM$_4$ as Solvable Matrix Quantum Mechanics,''
Nucl.Phys. {\bf B571} (2000) 479}
\lref\KZJ{V.A.~ Kazakov, P.~ Zinn-Justin,
``Two-Matrix model with ABAB interaction,''
Nucl.Phys. {\bf B546} (1999) 647.}
\lref\LS{R.G.~ Leigh, M.J.~ Strassler,
``Exactly Marginal Operators and Duality in Four Dimensional
$N=1$ Supersymmetric Gauge Theory,'' Nucl.Phys. {\bf B447} (1995) 95.}
\lref\FGZ{P.Di~ Francesco, P. ~Ginsparg, J. ~Zinn-Justin,
``2D Gravity and Random Matrices,'' Phys.Rept. {\bf 254} (1995) 1.}
\lref\Kazakov{V. A.~ Kazakov, ``Solvable Matrix Models,'' hep-th/0003064.}
\lref\Kostov{I.~Kostov, ``Exact Solution of the Six-Vertex Model
on a Random Lattice,'' Nucl.Phys. {\bf B575} (2000) 513.}
\lref\BDE{G.~Bonnet, F.~David, B.~Eynard,
``Breakdown of universality in multi-cut matrix models,''
J.Phys. {\bf A33} (2000) 6739.}
\lref\Dorey{N.~Dorey, T.~J.~Hollowood, S.~P.~Kumar and A.~Sinkovics,
``Massive vacua of N = 1* theory and S-duality from matrix models,''
hep-th/0209099.}
\lref\Superspace{S.J.~Gates Jr, M.T.~Grisaru, M.~Rocek, W.~Siegel,
``Superspace, or One thousand and one lessons in supersymmetry,'' 1983;
hep-th/0108200.}
\lref\witcs{E.~Witten, ``Chern-Simons gauge theory as a string theory,''
hep-th/9207094.}
\lref\Ginsparg{P.~Ginsparg
``Matrix models of 2d gravity,'' Trieste Lectures (1991), hep-th/9112013.}
\lref\dadda{A.~D'adda, ``Comments on Supersymmetric Vector and Matrix
Models,'' Class. Quant. Grav. {\bf 9} (1992) L21;
``New Methods of Integration in Matrix Models,''
Class. Quant. Grav. {\bf 9} (1992) L77.}
\lref\BF{ P.~F.~Byrd, M.~D.~Friedman, Handbook of elliptic integrals for engeneers and physicists, Springer-Verlag (1954).}
\lref\CVlast{F.~Cachazo, C.~Vafa,
``N=1 and N=2 Geometry from Fluxes,'' hep-th/0206017.}
\lref\SW{N.~Seiberg, E.~Witten, ``Monopole Condensation,
And Confinement In N=2 Supersymmetric Yang-Mills Theory,''
Nucl.Phys. {\bf B426} (1994) 19; Erratum-ibid. {\bf B430} (1994) 485.}
\lref\Klemm{A.~Klemm, W.~Lerche and S.~Theisen,
``Nonperturbative effective actions of N=2 supersymmetric gauge theories,''
Int.\ J.\ Mod.\ Phys.\ A {\bf 11} (1996) 1929, hep-th/9505150.}
\lref\Khoze{N.~Dorey, V.~V.~Khoze and M.~P.~Mattis,
``Multi-instanton check of the relation between
the prepotential F and  the modulus u in N = 2 SUSY Yang-Mills theory,''
Phys.\ Lett.\ B {\bf 390} (1997) 205, hep-th/9606199.}
\lref\Brezin{E.~Brezin, C.~Itzykson, G.~Parisi and J.~B.~Zuber,
``Planar Diagrams,'' Commun.\ Math.\ Phys.\  {\bf 59}, 35 (1978).}
\lref\GAKO{ I.~Kostov, `` The ADE face models on a fluctuating planar lattice'', 
Nucl.\ Phys.\ B {\bf 326} (1989) 583;    
M.~Gaudin and I.~Kostov, ``O(N) Model On A Fluctuating Planar Lattice:
Some Exact Results,'' Phys.\ Lett.\ B {\bf 220} (1989) 200.}
\lref\MMM{S.~Kharchev, A.~ Marshakov, A.~ Mironov, A.~ Morozov, S.~ Pakuliak,
``Conformal Matrix Models as an Alternative to Conventional Multi-Matrix
Models,'' Nucl. Phys. {\bf B404} (1993) 717, hep-th/9208044;
I.~K.~Kostov, ``Gauge invariant matrix model for the A-D-E closed strings,''
Phys.\ Lett.\ B {\bf 297} (1992) 74, hep-th/9208053.}

\lref\bbdo{
C.~Bachas, J.~de Boer, R.~Dijkgraaf and H.~Ooguri, ``Permeable
conformal walls and holography,'' JHEP {\bf 0206} (2002) 27,
hep-th/0111210.}
\lref\ov{
H.~Ooguri and C.~Vafa, ``Worldsheet derivation of a large $N$ duality,''
hep-th/0205297.}

\lref\BCOV{M.~Bershadsky, S.~Cecotti, H.~Ooguri and C.~Vafa,
``Kodaira-Spencer theory of gravity and exact results
for quantum string amplitudes,''
Commun.\ Math.\ Phys.\  {\bf 165} (1994) 311, hep-th/9309140.}
\lref\GVone{R.~Gopakumar and C.~Vafa,
``On the gauge theory/geometry correspondence,''
Adv.\ Theor.\ Math.\ Phys.\  {\bf 3} (1999) 1415, hep-th/9811131.}
\lref\Vafa{C.~Vafa, ``Superstrings and Topological Strings at Large N,''
hep-th/0008142.}
\lref\Doreyi{N.~Dorey, T.~J.~Hollowood, S.~Prem Kumar and A.~Sinkovics,
``Exact superpotentials from matrix models,'' hep-th/0209089.}
\lref\Aganagic{M.~Aganagic and C.~Vafa,
``Perturbative derivation of mirror symmetry,'' hep-th/0209138.}
\lref\Ferrari{F.~Ferrari,
``On exact superpotentials in confining vacua,'' hep-th/0210135.}
\lref\Fuji{H.~Fuji, Y.~Ookouchi,
``Comments on Effective Superpotentials via Matrix Models,''
hep-th/0210148.}
\lref\Berenstein{D.~Berenstein, ``Quantum moduli spaces from matrix
models,'' hep-th/0210183.}
\lref\Mironov{L.~Chekhov and A.~Mironov,
``Matrix models vs. Seiberg-Witten/Whitham theories,'' hep-th/0209085.}
\lref\ghov{D.~Ghoshal and C.~Vafa,
``C = 1 string as the topological theory of the conifold,''
Nucl.\ Phys.\ B {\bf 453} (1995) 121, hep-th/9506122.}
\lref\doet{N.~Dorey,
``An elliptic superpotential for softly broken N = 4
supersymmetric  Yang-Mills theory,''
JHEP {\bf 9907} (1999) 21, hep-th/9906011;
N.~Dorey and S.~P.~Kumar,
``Softly-broken N = 4 supersymmetry in the large-N limit,''
JHEP {\bf 0002} (2000) 6, hep-th/0001103.}
\lref\ADK{O.~Aharony, N.~Dorey and S.~P.~Kumar,
``New modular invariance in the N = 1* theory,
operator mixings and  supergravity singularities,''
JHEP {\bf 0006} (2000) 26, hep-th/0006008.}
\lref\DAVID{ F.~David, Non-Perturbative Effects in Matrix Models and Vacua of 
Two Dimensional Gravity, Phys.Lett. B302 (1993) 403-410; hep-th/9212106.}
\lref\KaMi{ V.A.~Kazakov and A.A.~Migdal, 
``Recent progress in the theory of noncritical strings'', Nucl.Phys. B
{\bf 311} (1988) 171.  }

\lref\ivan{
I.~K.~Kostov, ``Strings with discrete target space,'' Nucl.\ Phys.\ B
{\bf 376}, 539 (1992) [arXiv:hep-th/9112059].  }

%%%%%%%%%%%%%%%%%%% TITLE PAGE  %%%%%%%%%%%%%%%%%%%%%%%%%%%%%%%%

\Title{\vbox{\baselineskip11pt\hbox{hep-th/0210238}
\hbox{HUTP-02/A049}
\hbox{ITEP-TH-51/02}
\hbox{ITFA-2002-41}
}} {\vbox{
\centerline{Perturbative Analysis of Gauged Matrix Models}
\vskip 4pt
%\centerline{}
}}
\centerline{
Robbert Dijkgraaf,$^1$
Sergei Gukov,$^2$
Vladimir A. Kazakov,$^3$
and Cumrun Vafa$^2$}
%\medskip
%\medskip
%\medskip
\medskip
\vskip 8pt
\centerline{\it $^1$ Institute for Theoretical Physics \&}
\centerline{\it Korteweg-de Vries Institute for Mathematics,}
\centerline{\it University of Amsterdam,}
%\centerline{\it Valckenierstraat 65}
\centerline{\it 1018 XE Amsterdam, The Netherlands}
\medskip
\centerline{\it $^2$
Jefferson Physical Laboratory, Harvard University,}
\centerline{\it Cambridge, MA 02138, USA}
\medskip
\centerline{\it $^3$
Laboratoire de Physique Th\'eorique de l'Ecole Normale Sup\'erieure,}
\centerline{\it 24 rue Lhomond, 75231 Paris CEDEX, France}
\medskip
%\medskip
\medskip
%\bigskip
\noindent
We analyze perturbative aspects of gauged matrix models, including
those where classically the gauge symmetry is partially broken. Ghost
fields play a crucial role in the Feynman rules for these vacua. We
use this formalism to elucidate the fact that non-perturbative aspects
of $\CN=1$ gauge theories can be computed systematically using
perturbative techniques of matrix models, even if we do not possess an
exact solution for the matrix model. As examples we show how the
Seiberg-Witten solution for $\CN=2$ gauge theory, the Montonen-Olive
modular invariance for $\CN=1^*$, and the superpotential for the
Leigh-Strassler deformation of $\CN=4$ can be systematically computed
in perturbation theory of the matrix model/gauge theory (even though
in some of these cases the exact answer can also be obtained by
summing up planar diagrams of matrix models).

\smallskip
%\medskip
\Date{October 2002}

%%%%%%%%%%%%%%%%%%%%%%%%%%%%%%%%%%%%%%%%%%%%%%%%%%%%%%%%%%%%%%%%%%%%%

\newsec{Introduction}

In this paper we study perturbative aspects of matrix models as
applied to non-perturbative dynamics of $\CN=1$ supersymmetric gauge
theories in four dimensions (admitting a large $N$ description)
\refs{\DVone, \DVtwo, \DVthree}.  The connection between
the matrix model and the supersymmetric gauge theory proceeds by
identifying the superpotential of the gauge theory with the potential
of the matrix model.  It was shown in \refs{\DVone,\DVtwo,\DVthree},
building on previous work \refs{\BCOV,\GVone,\Vafa,\CIV}, that the
planar diagrams of the matrix model effectively compute the exact
glueball superpotential for the associated supersymmetric gauge theory
and thus yield, upon extremization, exact results for the gauge
theory.  There has been some further work in this direction
\refs{\Mironov, \Doreyi, \Dorey, \Aganagic, \Ferrari, \Fuji, \Berenstein}.

In some cases the planar diagrams of matrix model can be summed up
exactly.  This then gives rise to a dual geometry at the planar limit,
from which one can read off non-trivial holomorphic information about
the associated supersymmetric gauge theory.
In this respect it is interesting to note that
up to now all the cases where the
supersymmetric gauge theory can be solved using strong/weak coupling
dualities fall in the class of exactly soluble matrix models. In all
these cases the solution takes the form of a dual geometry. 
However, in most cases
({\it i.e.}\ for a generic matter content and interactions) the exact
solution of the corresponding matrix model is not available, even in
the planar limit.

But, even if the planar diagrams cannot be exactly summed, we still
can resort to perturbative techniques of the matrix model.  This
yields, as noted in \DVthree , a systematic instanton expansion in the
gauge theory.  Thus, for a large class of supersymmetric gauge
theories for which we had no dual descriptions, we can now
nevertheless compute in a systematic way instanton corrections to
interesting holomorphic quantities.  Thus, in a sense, we are going
beyond duality, and we may hope that this will ultimately give us a
new perspective about the meaning of duality in gauge theory and
string theory.

Perturbative techniques for matrix models are not completely trivial.
This is because we are dealing with a {\it gauged} matrix model, and
it is crucial to take this gauging into account properly.  For vacua
where the gauge symmetry is not broken, this can be easily taken into
account by dividing by the volume of the gauge group, which simply
leads to an overall factor.  However, for vacua where the gauge group
is partially broken, not only do we have to divide by the volume of
the unbroken gauge group, we also have to deal with naive flat
directions of the matrix fields, which are pure gauge degrees of
freedom.  To address this, we can implement the standard method of
Faddeev-Popov ghosts, now applied to the broken part of the gauge
group.  The main aim of this paper is to develop this further and
apply it to a number of interesting examples.  This will include
examples where we know the exact solutions as well as some where we do
not know how to sum up the planar diagrams.  Since our emphasis in
this paper is the applicability of perturbative techniques we
illustrate the power of the perturbation theory, even for some of the
examples where we do know how to sum up planar diagrams.  We will
consider in particular $\CN=1^*$ and Leigh-Strassler deformation of
the $\CN=4$ super-Yang-Mills, as well as $\CN=2$ Seiberg-Witten
geometry.

As a byproduct of the results of this paper, which might be
interesting to the matrix model specialists, we demonstrate how the
matrix models with several eigenvalue supports in the large $N$ limit
can be studied by means of the planar diagram technique and
established well-defined Feynman rules for it. (This subject is also
discussed in \BDE.) Another novelty which is not well explored in the
matrix model literature is the possibility of filling not only the
minima but also the maxima of the matrix potential (the ``unstable''
cuts), by virtue of the analytical continuation in the filling
parameters. We demonstrate this with the example of the one matrix
model with the cubic potential where we fill by eigenvalues both the
minimum and the maximum. One can show that this model is equivalent to
a particular case of the models of random paths studied in \GAKO,
where the solution can be written in terms of elliptic functions.

The organization of this paper is as follows: In section 2 we show how
gauge fixing in the one matrix model with the cubic potential is done,
when the classical vacuum partially breaks the gauge symmetry.  We
establish the planar diagrammatic rules for this model.  We show the
importance of ghosts for matrix models in this context and relate it
to the ghosts of the supersymmetric gauge theory. We also demonstrate
that the Feynman rules for the multi-cut solutions have a nice
geometric interpretation in terms of domain walls on the closed string
world-sheet. In section 3 we study various examples.  In appendix A we
recall how the exact solution can be obtained in the case of the cubic
superpotential as well as some connections with $c=1$ strings on the
self-dual radius.  In appendix B we show how to setup the perturbation
theory for massive vacua of $\CN=1^*$ where the rank of the gauge
group is reduced.

%\bigskip\bigskip\bigskip\bigskip
%\bigskip\bigskip\bigskip\bigskip

%%%%%%%%%%%%%%%%%%%%%%%%%%%%%%%%%%%%%%%%%%%%%%%%%%%%%%%%%%%%%%%%%%%%%

\newsec{Gauge Fixing in Field Theory and Matrix Models}

\subsec{The Problem}

\ifig\twoloop{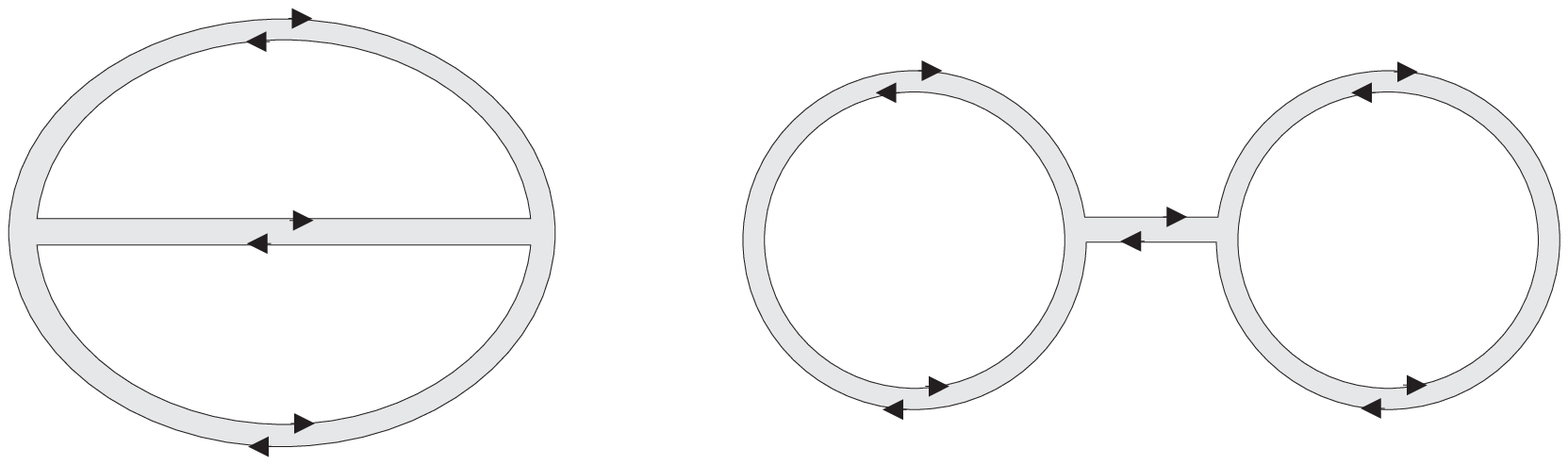}{100}{
The two planar two-loop diagrams, with combinatorial weight ${1\over
6}$ and ${1\over 2}$, that contribute to the order $S^3$ term in the
free energy.}

To explain the setup and review the proposal of
\refs{\DVone,\DVtwo,\DVthree}, let us start with a simple integral
over a single $M \times M$ matrix $\F$
\eqn\PARTF{
Z = {1\over {\rm vol}\,U(M)} \int d\F\,\exp
\left( {1\over g_s}\Tr\,W(\F) \right),  }
where $W(x)$ is a cubic polynomial with two critical points
at $x=a_1$ and $x=a_2$
\eqn\wcubic{W'(x)= (x-a_1)(x-a_2).}
It was explained in \DVone\ how to compute the genus zero free energy
in this model if we put all the eigenvalues of the matrix $\F$ at one
critical point, say at $a_1$.  Shifting the matrix as $\F \to a_1 {\bf
1}+\F$ we obtain (up to a constant)
\eqn\WWW{
W = \Tr \left( {1\over 2}\Delta \F^2+{1\over3} \F^3 \right)  }
with
$$
\Delta=a_1-a_2.
$$
From this action we easily read off the Feynman rules: a propagator
$1/\Delta$ for the $\Phi$ variable and a three-point vertex with
weight 1.  This gives for example the following two-loop contribution
to the perturbative part of the genus zero free energy, with
contributions ${1\over 6}$ and ${1\over 2}$ from the two planar
diagrams of \twoloop
\eqn\cfstwoloop{
\cF^{\rm pert}_0 = {2\over 3} {1\over \Delta^3} S^3 + \ldots}
Here $S=g_s M$ plays the role of the 't Hooft parameter.

According to \DVone, the planar limit of this matrix model can be used
to obtain exact holomorphic quantities in the corresponding $\CN=1$
gauge theory, which in this case is simply a $U(N)$ supersymmetric
gauge theory with a single adjoint superfield and a tree-level
superpotential $\Tr W(\F)$ given by \wcubic.  For example, the
effective superpotential is essentially given by the derivative of the
$\cF_0 (S)$,
\eqn\weffa{
W_{\rm eff} (S) = N S \log (S / \Lambda^3) - 2 \pi i \tau_0 S + N
{\partial \cF^{\rm pert}_0 (S) \over \partial S}}
where the first term can be seen as coming from the contribution of
the measure factor to the free energy $\cF_0$ \ov.
Here the variable $S$ is identified with the chiral glueball field,
$$
S = {1 \over 32 \pi^2} \tr \cW_{\alpha} \cW^{\alpha}.
$$
From the effective superpotential $W_{\rm eff}$ one can read off
non-perturbative information about the infra-red dynamics and vacuum
structure of $\CN=1$ theory.  Thus, critical points of $W_{\rm eff}$
generically correspond to massive vacua in the low-energy theory. On
the other hand, the difference $\Delta W_{\rm eff}$ between the value
of the superpotential at two different critical points determines the
tension of the BPS domain wall separating the two vacua.

In order to find the value of $W_{\rm eff}$ at each vacuum, one should
extremize it with respect to $S$ and then reexpress the result in
terms of the (bare) gauge coupling $\tau_0$.  As a result, one
typically finds an instanton expansion, in which the $n$-instanton
terms are fixed by the perturbative contributions to $\cF_0$ up to the
$n$-loop order.  For example, already the two-loop result \cfstwoloop\
can be used to determine $W_{\rm eff}$ {\it exactly} up to
two-instanton order.

It is important to stress here that the rank $M$ of the gauge group
in the matrix model is completely unrelated to the rank $N$ of the
gauge group in the corresponding $\CN=1$ theory.  In order to
appreciate this point, note that $M$ enters the effective
superpotential \weffa\ in a very complicated manner (via the $S$
dependence), whereas the $N$ dependence is very simple (linear).
In particular, the value of $N$ does not have to be large;
the result \weffa\ can be applied just as well to a $U(2)$ gauge theory.
Henceforth, we will be very careful to distinguish between $M$ and $N$.

Now let us proceed to a more general classical vacuum with $M_1$
eigenvalues at $a_1$ and $M_2$ eigenvalues at $a_2$
$$
\F = \left(\matrix{a_1 & 0 \cr 0 & a_2 \cr}\right).
$$
So in the matrix model we break the gauge symmetry as
$$
U(M) \to U(M_1) \times U(M_2).
$$
Within the string theory realization this corresponds to a background
with two clusters of D-branes of charge $M_1$ and $M_2$ respectively.
Taking both $M_1$ and $M_2$ to be large, we obtain a so-called two-cut
solution of the matrix model. To find the perturbative expansion of
this solution it is too naive to simply expand the matrix $\F$ around
this point. Indeed, if we shift
\eqn\shift{
\F \to \left(\matrix{a_1 & 0 \cr 0 & a_2 \cr}\right)+ \F,
}
and decompose the matrix $\F$ in blocks
\eqn\block{
\F=\left(\matrix{\F_{11} & \F_{12} \cr \F_{21} & \F_{22} \cr}\right)
}
(where $\F_{ij}$ corresponds to an $ij$ string, going from the
$i$th D-brane to the $j$th D-brane) then the quadratic piece in the
action takes the form
$$
{1\over 2}\Delta\, \Tr\(\F_{11}^2 + \F_{21}\F_{12} - \F_{12} \F_{21}
-\F_{22}^2 \)= {1\over 2}\Delta\, \Tr\(\F_{11}^2-\F_{22}^2\).
$$
So, the kinetic terms for the ``off-block diagonal'' components
$\F_{12}$ and $\F_{21}$ will vanish. This makes it problematic to keep
track of the 12 and 21 degrees of freedom.

This vanishing of the kinetic term for the off-diagonal components is
not surprising since they are zero-modes. The original $U(M)$ gauge
symmetry still acts on the matrix configurations and the broken gauge
transformations will transform a vacuum with two clusters of
eigenvalues into a gauge equivalent state. More precisely, we now have
a non-trivial vacuum manifold parametrized by the coset
$$
U(M)/U(M_1)\times U(M_2).
$$
Since the action is $U(M)$ invariant, the matrix integral will not
depend on the choice of point on this vacuum manifold. The
corresponding $2M_1M_2$ zero-modes are exactly the components
$\F_{12}$ and $\F_{21}$.

The correct way to treat the semi-classical expansion, keeping track
of the $M_1$ and $M_2$ dependence, is by the method of Faddeev-Popov
ghosts. We will see in a moment how this emerges both from the
four-dimensional gauge theory and from the matrix model. But let us
here remark that the role played by the ghosts is also suggested by
going back to the topological string derivation of the matrix model
as described in \DVone.  

There one starts from a reduction to two dimensions of six-dimensional
holomorphic Chern-Simons theory \witcs.  The six-dimensional open
string field theory contains fields of various ghost numbers that
correspond geometrically to differential forms of different degree on
the Calabi-Yau manifold. If we reduce the theory down to two
dimensions, we find at the physical ghost level (among other fields) a
gauged chiral scalar field $\F(z)$, whose zero-mode is the variable
$\F$ in the matrix integral.

But there is also a contribution of the ghosts in this two-dimensional
world-volume theory. One finds in particular a scalar ghost $C(z)$ and
a conjugate ghost $B(z)$, that is a $(1,1)$ form on the
world-volume. Both are adjoint valued, with action
$$
{1 \over g_s} \int d^2 z \, \Tr\(B\overline{D}_A C+ B[\F,C]\).
$$
Since both scalars $\F$ and $C$ reduce to their constant zero-modes,
only the overall volume factor in the two-form $B$ contributes in the
path-integral. So we get an additional ghost contribution to the
matrix integral of the form
\eqn\wgh{
W_{\rm ghost}= \Tr\( B[\F,C]\),
}
where $B,C$ are now anticommuting $M \times M$ matrices.  Let us now
explain in more detail the origin of this term more directly in the
four-dimensional $\CN=1$ gauge theory and in the corresponding matrix
model.

%%%%%%%%%%%%%%%%%%%%%%%%%%%%%%%%%%%%%%%%%%%%%%%%%%%%%%%%%%%%%%%%%%%%
\subsec{Gauge Fixing in $\CN=1$ Supersymmetric Gauge Theory}

Consider $\CN=1$ gauge theory with a $U(N)$ vector multiplet and one
chiral matter multiplet in the adjoint representation of the gauge
group. In $\CN=1$ superspace the field content of such theory is
represented by a vector superfield $V$ and an adjoint chiral scalar
superfield $\Phi$.  Let $S_{{\rm inv}} (V, \Phi, \bar \Phi)$ be the
action of the superfields $V$ and $\Phi$, invariant under $U(N)$ gauge
transformations
\eqn\gaugetransform{
e^{V} \to e^{i \bar \Lambda} e^V e^{- i \Lambda},}
where $\Lambda$ is a chiral gauge parameter.

Our goal will be to study (partial) gauge fixing in the functional integral
\eqn\pathint{
Z = \int \CD V \CD \Phi \CD \bar \Phi ~e^{S_{{\rm inv}} (V, \Phi, \bar \Phi)}}
by imposing a gauge fixing constraint on the adjoint scalar $\Phi$.
Implementing the standard Faddeev-Popov procedure, one finds: $(a)$
that (partial) fixing of the $U(N)$ gauge symmetry leads to new
anti-commuting chiral ghost superfields $B$ and $C$; and $(b)$ that
the ghost action can be written as an F-term of the form \wgh.

The first statement does not depend on the particular way of gauge
fixing.  It is simply related to the fact that the gauge parameter
$\Lambda$ is a chiral scalar and, therefore, the gauge-fixing function
$F=F(V,\Phi, \bar \Phi)$ should also be a chiral superfield.  namely,
the gauge constraint should be of the form \Superspace:
\eqn\gaugefix{F = f, \quad \bar F = \bar f}
where $f=f(x,\theta)$ is some chiral function.  As we review below,
this implies that the ghost superfields are also chiral.

On the other hand, the second statement above relies on the assumption
that the gauge-fixing function $F$ does not depend on the vector
superfield $V$. Since, as we just explained, $F$ has to be chiral we
conclude that $F=F(\Phi)$.  In particular, a convenient choice of
gauge is given by a linear function $F(\Phi)$. Then, it follows from
the gauge transformation of $\Phi$, that under $U(N)$ gauge symmetry
$F$ transforms as:
$$
\delta F = [\Phi, \Lambda]
$$

Now, in order to apply the usual Faddeev-Popov method to the gauge
condition \gaugefix, we introduce the functional determinant:
$$
\Delta_F = \int \CD \Lambda \CD \bar \Lambda~
\delta (F - f) ~\delta (\bar F - \bar f)
$$
Inserting 1 into the path integral \pathint\ in the form
$\Delta_F \Delta_F^{-1}$, we obtain
$$
Z = \int \CD V \CD \Phi \CD \bar \Phi
~\Delta_F^{-1} ~\delta (F-f) ~\delta (\bar F - \bar f)
~e^{S_{{\rm inv}} (V, \Phi, \bar \Phi)}
$$
Introducing the chiral ghost fields $B$, $C$ and 
expressing the Faddeev-Popov determinant
$\Delta_F^{-1}$ in terms of the ghost action:
$$
\eqalign{
\Delta_F^{-1} & =
 \int \CD B \CD \bar B \CD C \CD \bar C~
\exp \Big[
\tr \int d^4 x d^2 \theta B
\left( {\delta F \over \delta \Lambda} C
+ {\delta F \over \delta \bar \Lambda} \bar C \right)\cr
& \qquad \qquad
+ \tr \int d^4 x d^2 \bar \theta ~\bar B
\left( {\delta \bar F \over \delta \Lambda} C \
+ {\delta \bar F \over \delta \bar \Lambda} \bar C \right)
\Big] \cr
& = \int \CD B \CD \bar B \CD C \CD \bar C~
\exp \Big[ \tr \int d^4 x d^2 \theta B [\Phi, C] + c.c. \Big] = \cr
& = \int \CD B \CD \bar B \CD C \CD \bar C~ e^{S_{{\rm ghost}}}
}
$$
leads to the path integral
\eqn\zghost{
Z = \int \CD V \CD \Phi \CD \bar \Phi \CD B \CD \bar B \CD C \CD \bar C
~e^{S_{{\rm inv}} + S_{{\rm GF}} + S_{{\rm ghost}}}  }
%
%Since, by construction, this integral does not depend on $f$ we can average
%over the latter with any measure. In particular, we can choose
%the measure to be Gaussian
%$$
%\int \CD f \CD \bar f ~\exp \left(
%- {1 \over \a} \tr \int d^4 x d^4 \theta ~ \bar f f \right)
%$$
%This
%This leads to the path integral
%%
%\eqn\zghost{\eqalign{
%Z & = \int \CD V \CD \Phi \CD \bar \Phi
%~\Delta_F^{-1} ~e^{S_{{\rm inv}} + S_{{\rm GF}}} \cr
%& = \int \CD V \CD \Phi \CD \bar \Phi \CD B \CD \bar B \CD C \CD \bar C
%~e^{S_{{\rm inv}} + S_{{\rm GF}} + S_{{\rm ghost}}}
%}}
%
%where the gauge-fixing term
%$$
%S_{{\rm GF}} = - {1 \over \a} \tr \int d^4 x d^4 \theta~ \bar F F
%= - {1 \over \a} \tr \int d^4 x d^4 \theta~ \bar \Phi \Phi
%$$
%simply renormalizes the kinetic term for $\Phi$.  Also, in the last
%line of \zghost\ we expressed the Faddeev-Popov determinant
%$\Delta_F^{-1}$ in terms of the ghost action:
%$$
%\eqalign{
%\Delta_F^{-1} & =
% \int \CD B \CD \bar B \CD C \CD \bar C~
%\exp \Big[
%\tr \int d^4 x d^2 \theta B
%\left( {\delta F \over \delta \Lambda} C
%+ {\delta F \over \delta \bar \Lambda} \bar C \right)\cr
%& \qquad \qquad
%+ \tr \int d^4 x d^2 \bar \theta ~\bar B
%\left( {\delta \bar F \over \delta \Lambda} C \
%+ {\delta \bar F \over \delta \bar \Lambda} \bar C \right)
%\Big] \cr
%& = \int \CD B \CD \bar B \CD C \CD \bar C~
%\exp \Big[ \tr \int d^4 x d^2 \theta B [\Phi, C] + c.c. \Big] = \cr
%& = \int \CD B \CD \bar B \CD C \CD \bar C~ e^{S_{{\rm ghost}}}
%}
%$$
where $S_{{\rm GF}}$ is the gauge-fixing action
and $S_{{\rm ghost}}$ is given by
$$
S_{{\rm ghost}} = \int d^4 x d^2 \theta ~ \tr \( B [\Phi, C] \) + c.c.
$$
This is the tree-level contribution to the superpotential that we were
after.  Specifically, it shows that for a (partial) gauge fixing via
imposing constraints on the adjoint chiral superfield $\Phi$, the
ghost action can indeed be written as the F-term.  Moreover, the form
of this term is exactly the same as the form of the ghost term \wgh\
in the matrix model action, which is in line with the general
statement that potential in matrix model should be identified with
classical superpotential in $\CN=1$ gauge theory \DVone.

%%%%%%%%%%%%%%%%%%%%%%%%%%%%%%%%%%%%%%%%%%%%%%%%%%%%%%%%%%%%%%%%%%%%
\subsec{Gauge Fixing in Matrix Models}

The ghost term \wgh\ can also be derived directly in the matrix model
by gauge fixing the $U(M)$ gauge symmetry that acts by conjugation on
$\F$
$$
\F \to U \cdot \F \cdot U\inv.
$$
A convenient gauge choice is putting $\F$ to diagonal form. This gives
the condition
$$
\F_{ij}=0,\qquad i\not=i.
$$
Implementing this gauge fixing through the BRST formalism introduces
exactly the above ghost fields; see \refs{\Ginsparg,\dadda} for more
discussion of ghost fields and gauge fixing in matrix models.

Decomposing the ghosts also in the block form \block, we see that
after the shift \shift\ the kinetic term of the ghosts is given by
$$
\Delta \, \Tr\(B_{21} C_{12}\) - \Delta\, \Tr\(B_{12} C_{21}\).
$$
So, in the case of the ghosts it is the 11 and 22 blocks that are not
propagating and the 12 and 21 block that are ``physical.''

We conclude that in the reduction to the matrix integral the 11 and 22
strings represent physical matter fields and that the 12 and 21
strings represent ghost degrees of freedom. This makes sense
physically, since, as we already explained, in this two-cut classical
vacuum with reduced gauge symmetry $U(M_1) \times U(M_2)$ the matrix
elements in the 11 and 22 blocks cannot be obtained by gauge
transformations and thus they are classically not pure gauge, whereas
the 12 and 21 blocks are pure gauge. In perturbation theory we
therefore are left with only the ghosts in the 12 and 21 blocks.

%In the language of the four-dimensional gauge theory this is just the
%Higgs mechanism at work. The GSO projection in the string theory
%correlates spins and ghost number in such a way that the zero forms in
%the CS theory reduce to gauge fields in four dimensions, and the one
%forms to scalars. Thus the ghosts $B,C$ in the matrix model correspond
%to four-dimensional gauge fields, and the field $\F$ (that was
%originally a component of the gauge field in the CS theory)
%corresponds to the Higgs scalar field. So, although the 12 and 21
%components of $\F$ become massless (zero quadratic term) this is
%harmless, since these unphysical modes are eaten up by the gauge field
%(ghosts) which becomes massive. Note that the mass $\Delta=a_1 - a_2$
%is exactly what the Higgs mechanism is supposed to give.

Before we turn to the Feynman rules that all this implies, let us
point out that this interpretation is consistent with the multi-cut
solution of the large $M$ limit of the matrix integral. Here we first
reduce the matrix integral to eigenvalues
\eqn\EVL{
Z = \int \prod_I d\l_I \prod_{I<J}\(\l_I-\l_J\)^2 \exp {1\over g_s}
\sum_I W(\l_I).  }
In the case of a two-cut solution we can split the eigenvalues $\l_I$
in two subsets. The first subset of $M_1$ eigenvalues $\l_I^{(1)}$ are
located around the first critical point $a_1$, the second subset of
$M_2$ eigenvalues $\l_J^{(2)}$ are located around the second critical
point $a_2$. In a semi-classical expansion these two critical points
and the corresponding eigenvalues can be thought to be well-separated.
We can regard the two sets $\{\l_I^{(1)}\}$ and $\{\l_J^{(2)}\}$ as
eigenvalues of two matrices, a $M_1 \times M_1$ matrix $\F_{11}$ and a
$M_2 \times M_2$ matrix $\F_{22}$ with matching potentials $W$.  In
the saddle-point approximation after the shift
\shift\ this gives the action
\eqn\xypot{
W_{\rm tree} = \Tr\({1\over 2}\Delta \F_{11}^2 +{1\over3} \F_{11}^3\)
+\Tr\(-{1\over 2}\Delta \F_{22}^2+{1\over3} \F_{22}^3\).
}

{}From the eigenvalue representation of the matrix integral it is clear
that the only way these matrices $\F_{11}$ and $\F_{22}$ interact is
through the Jacobian factor
$$
\prod_{I,J} \(\l_I^{(1)} - \l_J^{(2)}\)^2.
$$
(This is clearly true for arbitrary $W$.) This term can be
exponentiated directly in the action (see also \BDE) giving the
effective action
$$
2 \Tr\log\left(\F_{11}\otimes {\bf 1} - {\bf 1} \otimes \F_{22}\right).
$$

To bring out clearly the $M_1$ and $M_2$ dependence, this part of the
Vandermonde determinant can also be exponentiated by using the two
pairs of ghosts $(B_{21},C_{12})$ and $(B_{12},C_{21})$. (We have two
pairs because of the square of the Vandermonde in \EVL.) In order to
reproduce the right determinant the action of these ghosts should be
\eqn\wghost{
\eqalign{
W_{\rm ghost}=  &\Tr \(B_{21} \F_{11} C_{12} + C_{21}\F_{11} B_{12}\) \cr
& \qquad
+ \Tr \(B_{12} \F_{22} C_{21} + C_{12}\F_{22} B_{21}\).\cr}
}
But this is exactly the action \wgh\ restricted to the propagating
fields: the 11 and 22 blocks of $\F$ and the 12 and 21 blocks of
$B,C$.

{}From the two contributions to the action \xypot\ and \wghost\ we can
read off the Feynman rules.  We have propagators (we suppress the
obvious matrix indices)
$$
\eqalign{
\<\F_{11} \F_{11}\> & =  {1\over \Delta}, \cr
\<\F_{22} \F_{22}\> & =  -{1\over \Delta}, \cr
\<B_{12} C_{21}\> & = \< B_{21} C_{12}\> = {1\over \Delta},\cr
}
$$
and all three-point vertices have weight 1.

\ifig\twocuts{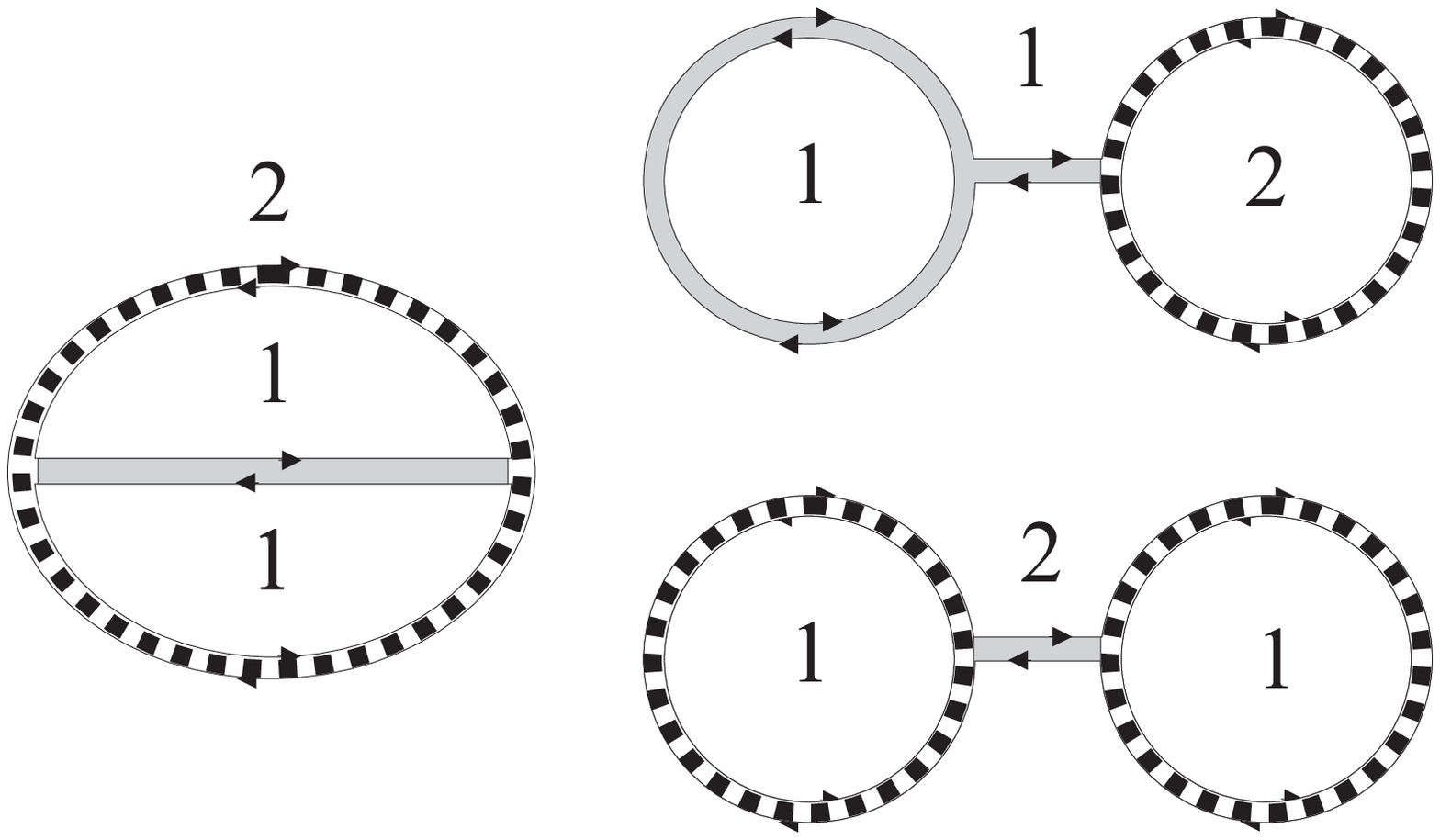}{80}{
The three planar two-loop diagrams, with combinatorial weight ${1\over
2}$, $1$ and ${1\over 2}$ respectively, that contribute to the order
$S_1^2S_2$ term in the free energy. The grey propagator indicates a
bosonic $\F_{11}$ or $\F_{22}$ field; the dashed propagator indicates
a $B,C$ ghost of type 12 or 21. The labeling of the hole or index loop
is also indicated.}

As a check of this perturbative prescription with the known properties
of the two-cut solution we will compute in this case the two-loop
contribution to the free energy $\cF_0(S_1,S_2)$. From the explicit
answer to the large $M$ solution we know this term is given by \CIV:
\eqn\TWOLOOP{
{1\over \Delta^3} \left({2\over 3}S_1^3 - 5 S_1^2 S_2 + 5 S_1 S_2^2
-{2\over 3} S_2^3\right)   }
The coefficients $\pm 2/3$ have already been computed. They come from
the two diagrams in \twoloop\ in which only $\F_{11}$ and $\F_{22}$
(and no ghosts) propagate.

The coefficients $\pm 5$ are given by the mixed diagrams in which also
the ghosts $B,C$ appear. Now there are three diagrams to consider,
which are given in \twocuts. Here the following factors contribute to
the weight of the diagram: the symmetry factor of the (colored) graph,
the extra minus signs of the ghost loops, the extra minus sign for the
$\F_{22}$ propagator compared to the $\F_{11}$ propagator, and the
fact that their are two flavors of ghosts ($B$ and $C$) running
through each ghost loop. With these considerations taken into account,
the three diagrams give a total combinatorial weight to the $S_1^2
S_2$ term of
$$
{1\over 2} \cdot (-1) \cdot 2 + 1\cdot (-1) \cdot 2 + {1\over 2}
\cdot (-1)^2 \cdot (-1) \cdot 4 = - 5.
$$
This indeed reproduces the second and third term in \TWOLOOP.

%{\bf The same in my notations. May be omited later  }
%
%We can regroup the ghosts in a two component complex $N_1\times N_2$
%matrix field $D_i,\ i=1,2$ in such a way that the action of model
%\xypot,\wghost\ would look as follows
%%
%\eqn\XWD{ \eqalign{
%W(X,Y,D)&= \Tr \(\hf X^2 +{g\over3} X^3\ -\hf
% Y^2\ +{1\over3} Y^3\cr
%&+ D_i^\dagger D_i+g\ D_i^\dagger D_i X - g D_i D_i^\dagger Y \). }}
%%
%where we have put $\Delta=1$ and rescaled $X,Y\to g X, g Y$, where
%$g^2=g_s$.  Note that the kinetic term for the ghost field
%$D_i^\dagger D_i$ appeared due to the shift \shift . It allows to give
%the following combinatorial interpretation of this model presented on
%the fig.2: a (planar) Feynman graph consists of the solid double
%lines of the field $A$ (propagator $= -1$), of the dotted double lines
%of the field $B$ (propagator $= +1$) and of the mixed (one solid, one
%dotted line) double lines of the field $C$ (propagator $=+1$). The
%triple vertices $X^3$, $Y^3$ and $DD^\dagger Y$ have the weight
%$(+g)$, whereas the vertex $D^\dagger DX$ has the weight $(-g)$. Due
%to the $O(4)$ symmetry of the combinations of $D$-fields $D_i^\dagger
%D_i$ or $D_iD_i^\dagger$, the propagators of $D$ form closed loops
%each separating the solid line ``phase'' from the dotted line
%``phase''. Each loop enters with the factor $(-2)$.  The expansion
%goes in powers of $g^2$ only, as usual for the cubic graphs.

%%%%%%%%%%%%%%%%%%%%%%%%%%%%%%%%%%%%%%%%%%%%%%%%%%%%%%%%%%
\subsec{ Relation to $\hat A_2$ and  $O(2)$ models  on planar graphs }

We will now argue that this two-cut model corresponds to the $\hat
A_2$ ``quiver'' model\foot{The corresponding Coxeter diagram consists
of a circle with two nodes.} on planar graphs introduced and studied
in \MMM.  Indeed, let us consider the Feynman rules of the previous
subsection (we choose the dimensionful parameter $\Delta=1$): if we
revert at the same time the sign of the propagator
$\langle\Phi_{22}\Phi_{22}\rangle$ from $+1$ to $-1$ and the sign of
the weight of each ghost loop from $+2$ to $-2$, it is the same as to
revert the sign of $S_2$. The latter will lead to only positive
coefficients in the formulas of the type \TWOLOOP\ for the expansion
for $\CF_{0}(S_1,S_2)$ given in the next section.  It is easy to check
this statement inductively: if we add one
$\langle\Phi_{22}\Phi_{22}\rangle$ to any diagram (like diagrams in
fig. 2) it adds up one extra loop weighted with the factor $S_2$, so
their sign changes are compensated. The same about a ghost loop: its
addition leads to a new loop with the $S_2$ factor, so their sign
changes are again compensated.
 
Hence we can write down the equivalent matrix model with the
potential:
$$
W= \Tr\[
\hf\Phi_{1}^2+{1\over 3}\Phi_{1}^3+\hf\Phi_{2}^2+{1\over 3}\Phi_{2}^3+
\hf \BC^\dagger \BC+ \BC^\dagger \BC\Phi_{1}+ \BC\BC^\dagger\Phi_{2} \],
$$
where $\Phi_1$ and $\Phi_2$ are $M_1\times M_1$ and $M_2\times M_2$
matrices, respectively, and $\BC=(C_1,C_2)$ is a vector of two
$M_1\times M_2$ rectangular complex matrix {\it bosonic} ghosts.  We
recognize here actually the $\hat A_2$ ``quiver'' matrix model with a
specific matrix potential.

In the symmetric case $S=S_1=-S_2$ this model is equivalent ({\it only
in the planar limit}, the difference due to the uncontractible ghost
loops on graphs of a nontrivial topology) to the $O(2)$ model
describing the statistics of selfavoiding (ghost) loops on planar
$\Phi^3$ type graphs, with the factor +2 for each loop (in the more
general $O(n)$ model one has the weight $n$ for each loop
\refs{\GAKO,\ivan}). This model is known to describe 2D 
quantum gravity coupled to the $c=1$ matter at the selfdual
compactification radius. In Appendix A we review the full planar
solution \CIV\ of this model from the one matrix model setup. In the
symmetric case the result is presented in terms of elliptic
parametrization.

\subsec{Multiple phases and domain walls on the world-sheet}

\ifig\resolve{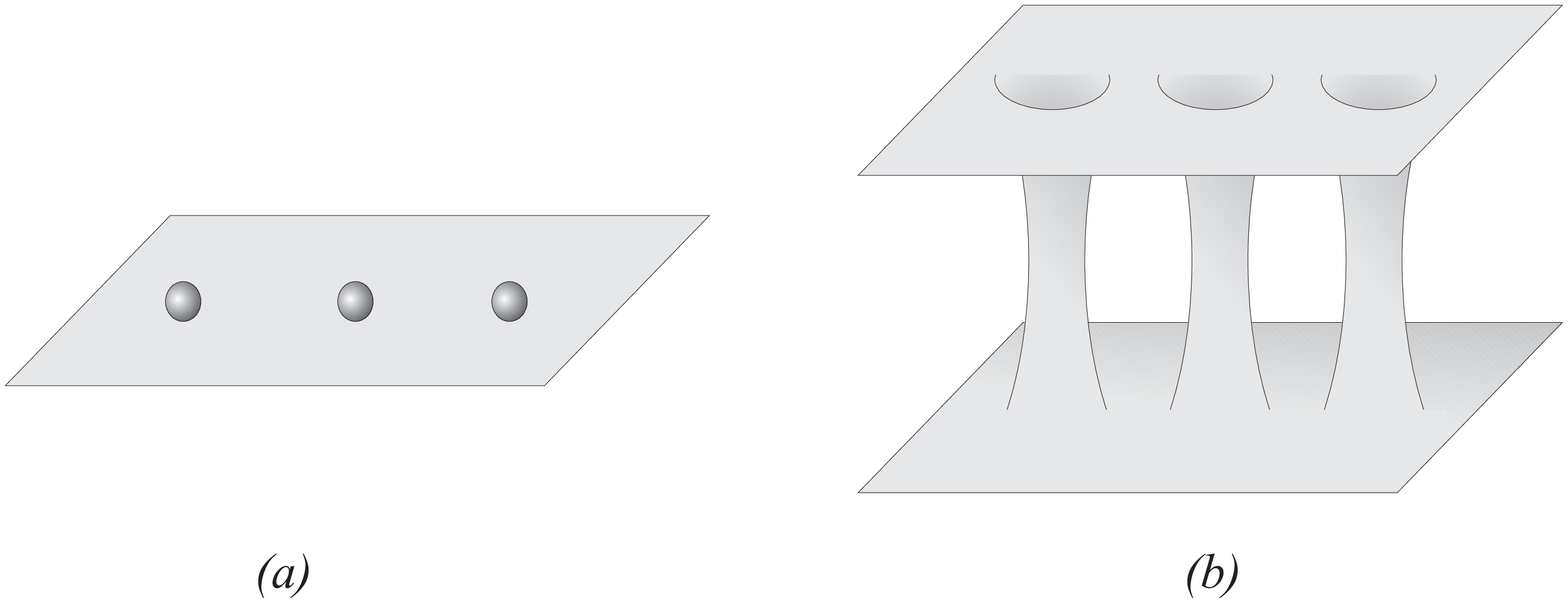}{120}{(a) The distribution of eigenvalues
at $g_s=0$; (b) The dual geometry (spectral curve) at finite 't Hooft
coupling.}

We would like to put the above construction into a bit more general
perspective. As we already mentioned we are dealing with a toy model
for a brane configuration where we have well-separated clusters of
$M_1,M_2,\ldots$ D-branes. In our toy matrix model we can see clearly
how such a multi-center geometry looks like from the open and closed
string perspectives. This might be helpful for understanding
gauge/gravity dualities for these kind of configurations in general.

In the matrix model at zero coupling ($g_s=0$) such a vacuum state is
simply given by the distribution of the eigenvalues in groups over the
critical points of $W$ in the complex eigenvalue plane as sketched in
\resolve(a). The eigenvalue density is represented as a sum of
delta-functions.

With the use of the large $N$ matrix model techniques we know that in
the dual closed string picture this geometry gets modified at non-zero
't Hooft coupling \DVone. The continuous eigenvalue density spreads
out along branch cuts in the eigenvalue plane. In this way a
non-trivial CY geometry emerges that is essentially given by a
hyperelliptic curve obtained as a double cover of the eigenvalue plane
as sketched in \resolve(b):
\eqn\geometry{y^2 = W'(x)^2 + {\rm deformations}}

Intuitively the following happens: if we insert a large number of
eigenvalues $M_i$ at the $i$th critical point of $W$ this builds up a
throat region in the dual geometry where the circumference of the neck
is given by the 't Hooft coupling $g_sM_i$. This fact that the size of
the geometry is proportional to the rank of the matrix, which is a
measure of the total number of degrees of freedom, should be thought
of as a version of the Bekenstein-Hawking geometric entropy, and it
would be interesting to develop this interpretation further.

We have seen that in the open string picture the character of the $ij$
strings, stretching from the $i$th to the $j$th D-brane, is very
different depending on whether $j=i$ or $j\not=i$. The diagonal $ii$
strings have interactions among themselves that are given by the
expansion of the superpotential $W$ around the $i$th critical point
and can therefore be of arbitrary order. These interactions build up
the fishnet double-line Feynman diagrams that in the large $N$ limit
will describe the closed string world-sheet propagating in the local
geometry around the $i$th D-brane, just as in the case of a single
center geometry.

\ifig\twophase{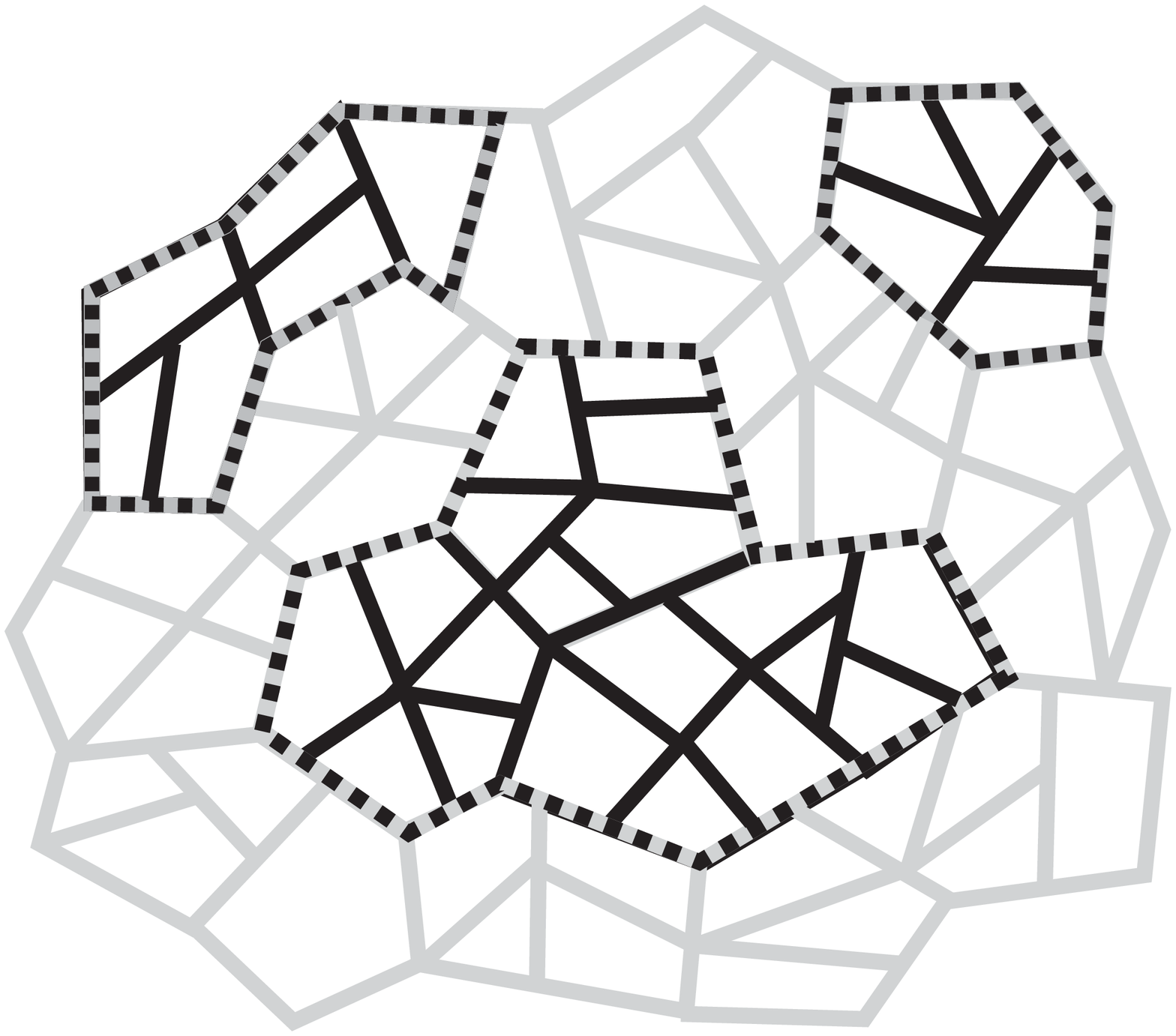}{60}
{In a two-cut solution the 11 strings and 22 strings (here indicated
in grey and black) will build up world-sheet theories out of fishnet
diagrams with interaction given by the (super)potential expanded
around the relevant critical point. The 12 and 21 strings (here
indicated by dashed lines) form self-avoiding loops, separating the
two phases on the world-sheet.}

The interactions of the off-diagonal $ij$ strings with $j\not=i$ do
{\it not} depend on the potential $W(\F)$. They are given entirely by
the cubic interaction \wghost\ that is dictated by gauge
invariance. Note that the action is quadratic in these $i\not=j$
strings --- ghost number is conserved --- and therefore the ghost
loops will form well-defined demarcation lines on the closed string
world-sheet separating the ``phase'' where the string is propagating
in the background of the $i$th D-brane from the phase where the string
propagates in the background of the $j$th D-brane, as sketched in
\twophase.  Because the absence of interactions among the $ij$ strings
these loops are self-avoiding.

In this way we observe that the multi-cut solutions of the matrix
model translated into a closed string picture naturally describe a
system of dynamical domain walls on the world-sheet. These domain
walls connect different conformal field theories as was analyzed in
\bbdo. In the open string channel the domain wall corresponds to an
$ij$ string stretching from one throat to another. This picture of
different phases of the world-sheet of a single closed topological
string is a further application of the ideas in \ov.

%%%%%%%%%%%%%%

\newsec{Examples}

In this section we illustrate how matrix perturbation theory can be
used to obtain non-perturbative instanton effects in various
supersymmetric gauge theories.  We start with some familiar examples,
which include $\CN=2$ Seiberg-Witten theory and $\CN=1^*$ theory,
where the exact answer is known to all orders.  Despite the existence
of the exact solution in these models, we will not need it
here. Instead, our goal is to reproduce it by computing simple planar
diagrams in the corresponding matrix model.

Of course, the real power of the perturbative technique is in those
models where exact solution is not available.  It is easy to come up
with simple examples of such models.  A particular example that we
discuss in this section is a massive deformation of the
Leigh-Strassler theory, which in turn is an (exactly marginal)
deformation of the $\CN=4$ super-Yang-Mills \LS.  The case that we
consider corresponds to a simple 3-matrix model with cubic
interactions, solution to which is not known even in the planar
limit. Nevertheless, one can systematically obtain instanton
corrections to the effective superpotential from matrix perturbation
theory.  Similar perturbative analysis can be applied essentially
to any $\CN=1$ theory that admits a large $N$ limit.

\subsec{Seiberg-Witten Solution from Multi-Cut Matrix Models}

The fact that one can obtain the Seiberg-Witten solution from a
perturbative analysis of the gauge theory, which in turn gets reduced
to planar computations of a matrix model has already been noted in
\DVthree\ as an interpretation of the string inspired derivation of
Seiberg-Witten geometry in \CVlast .  Our aim in this section is to
show that even if the exact solution of matrix model were not
available we could have nevertheless obtained a systematic instanton
expansion for quantities of interest. So in this section we are tying
one hand behind our back.

The basic idea of \CVlast\ is to deform $\CN=2$ theory to $\CN=1$ by a
polynomial tree-level superpotential $W(x)$, which freezes the
eigenvalues of the adjoint field $\Phi$ to a particular point on the
Coulomb branch.  For example, in the case of $U(2)$ gauge theory one
deforms by a cubic superpotential of the form \wcubic:
$$
W_{\rm tree}' (x) = \epsilon (x-a)(x+a).
$$
Here we explicitly introduced the deformation parameter $\epsilon$,
such that $\epsilon=0$ corresponds to the undeformed $\CN=2$ theory.
Choosing the configuration where one eigenvalue of $\Phi$ is at $+a$
and the other is at $-a$ determines a point on the Coulomb branch of
the original $\CN=2$ theory, and breaks the gauge group to an abelian
subgroup,
$$
U(2) \to U(1) \times U(1).
$$

This leads us precisely to the situation discussed in the previous
section, where we studied vacua of $\CN=1$ field theories with
(partial) gauge symmetry breaking.  Therefore, one should be able to
compute all holomorphic quantities from the genus zero free energy
$\CF_0 (S_1,S_2)$ of the corresponding two-cut matrix
model. Evaluating the two-loop Feynman diagrams in the previous
section we found the leading perturbative behaviour of the genus zero
free energy in the two-cut matrix model with a cubic interaction:
$$
\CF^{\rm pert}_{0} (S_1, S_2) =
{1\over \Delta^3} \left({2\over 3}S_1^3 - 5 S_1^2 S_2 + 5 S_1 S_2^2
-{2\over 3} S_2^3\right) + \ldots
$$
One can go further and systematically compute higher-order
corrections. In this way one finds a series expansion
\eqn\fpertt{\eqalign{
& \CF_{0} (S_1, S_2) =
- {1 \over 2} \sum_{i=1,2} S_i^2 \log \left( {S_i \over \Delta^3} \right)
+ (S_1 + S_2)^2 \log \left( {\Lambda \over \Delta} \right)
+  \cr
&~~~~ + {1\over \Delta^3} \left(
{2\over 3}S_1^3 - 5 S_1^2 S_2 + 5 S_1 S_2^2 - {2\over 3} S_2^3
\right) + \cr
&~~~~ + {1\over \Delta^6} \left(
{8 \over 3} S_1^4 - {91 \over 3} S_1^3 S_2 + 59 S_1^2 S_2^2
- {91 \over 3} S_1 S_2^3 + {8 \over 3} S_2^4
\right) + \cr
&~~~~ + {1\over \Delta^9} \left(
{56 \over 3} S_1^5 - {871 \over 3} S_1^4 S_2
+ {2636 \over 3} S_1^3 S_2^2 - {2636 \over 3} S_1^2 S_2^3
+ {871 \over 3} S_1 S_2^4 - {56 \over 3} S_2^5
\right) + \ldots
}}
Here the first term receives a contribution from the measure of the unbroken
gauge group $U(M_1) \times U(M_2)$ \ov, where each factor gives a standard
term $S_i^2 \log S_i$ that reproduces the Veneziano-Yankielowicz
superpotential. The one-loop diagrams for $\F$ and the ghosts $B,C$ 
account for the $\Delta$ dependence of the first two terms in \fpertt\
$$
\left({1\over 2} S_1^2 -2 S_1 S_2 + {1\over 2} S_2^2\right) \log \Delta.
$$
Finally, the $\Lambda$ dependence reflects the ambiguity in the cut-off of
the full $U(M_1+M_2)$ gauge group and should therefore multiply
$(S_1+S_2)^2$. The higher order perturbative terms have the combinatorial
meaning we explained in the previous section. For example, the terms that 
involve only $S_1$ or $S_2$ enumerate planar cubic diagrams and were 
computed in \Brezin.

Note that the function $\cF_0(S_1,S_2)$ is symmetric in $S_1$ and
$-S_2$. This reflects the symmetry of the potential: we can exchange
the stable and unstable critical points if we change the overall sign
of the potential by $g_s \to -g_s$. Since $S_i=g_sM_i$ this gives $S_1
\leftrightarrow -S_2$. From the combinatorial point of view this was
explained in section 2.4 in terms with the connection to the $O(2)$
model on a random surface --- it is an obvious property of the Feynman
rules.

We should now extremize the effective glueball superpotential
\eqn\weffsw{
W_{\rm eff} (S) = \sum_i \Big(
% N_i S_i \log (S_i / \Lambda_i^3) +
N_i {\partial \CF_{0} (S) \over \partial S_i}
- 2 \pi i \tau_0 S_i \Big) }
In the present case we have $N_1 = N_2 = 1$ and we will also set to
zero the bare coupling $\tau_0$. 

The physical quantity to compute in this model is the matrix of the
$U(1)\times U(1)$ couplings in the effective low-energy theory. It is
given by the second derivatives of matrix model free energy
\eqn\tauij{
\tau_{ij} = {\partial^2  \CF_{0} (S) \over \partial S_i \partial S_j }.
}
Note that by a scaling argument the matrix $\tau_{ij}$ does not depend
on the deformation parameter $\epsilon$ and therefore it should
reproduce the coupling constant of the $\CN=2$ Seiberg-Witten theory
at the relevant point of the Coulomb branch. Minimizing the effective
superpotential \weffsw, that in this case simplifies to
$$
W_{\rm eff}(S)= \sum_i {\partial \CF_{0} (S) \over \partial S_i},
$$
gives the condition
$$
\sum_i \tau_{ij}=0.
$$
So we see that at the extremum $\tau_{ij}$ takes the form
\eqn\ttt{
\pmatrix{ \tau_{11} & \tau_{12} \cr
\tau_{21} & \tau_{22}} 
= \tau \pmatrix{ 1 & -1 \cr -1 & 1},
}
where $\tau$ is the effective gauge coupling for the `off-diagonal'
$U(1) \subset SU(2) \subset U(2)$.  Note that we automatically managed
to get rid of the diagonal $U(1)$ factor by setting the bare coupling
constant to zero in eq. \weffsw.

The extremization of $W_{\rm eff}(S)$ we can do using the perturbative
expansion of $\cF_0$ \fpertt. However, before we do this, let us
recall that, in terms of the exact solutions, this extremization has a
clear geometric interpretation \refs{\CIV,\CVlast}. The free energy
$\cF_0$ can be described in terms of the dual geometry \geometry\ that
in this case of a cubic superpotential takes the form of a genus one
curve
\eqn\curve{
y^2=(x^2-a^2)^2+ b_1x + b_0.
}
Here the coefficients $b_1,b_0$ are determined by the 't Hooft
couplings $S_1,S_2$. In particular one has the simple relation $b_1=
-4(S_1+S_2)$. Minimizing $W_{\rm eff}(S)$ with respect to $S_1$ and
$S_2$ gives the condition
\eqn\sss{
S_1=-S_2.}
Therefore the algebraic curve \curve\ reduces to nothing but the
Seiberg-Witten curve for $SU(2)$ theory \SW:
$$
y^2 = (x^2 - u)^2 + \Lambda^4,
$$
where one has to make the identification of parameters (with
$\Delta=2a$)
\eqn\udelrel{
u ={1 \over 2} \langle \tr \Phi^2 \rangle = {1 \over 4} \Delta^2.
}
So at the extremum the free energy $\cF_0$ can be thought of as a
function of only one variable $S=S_1=-S_2$ that is determined by the
parameter $\Delta$ (or $u$) of the SW curve. 

Both in the matrix model and in the SW solution the exact expression
for the coupling $\tau$ of the off-diagonal $U(1)$ that appears in
\ttt\ follows directly from this geometric interpretation as the
modulus of an elliptic curve.  Given the parametrization of this
curve, we can expand $\tau$ in terms of the variable $u$ or $\Delta$
and obtain the exact result
\eqn\matrixtau{
\tau(u)  = 2 \log \left( {\Lambda \over \Delta} \right)
+ 20 \left( {\Lambda \over \Delta} \right)^4 + 538 \left( {\Lambda
\over \Delta} \right)^8 + {62048 \over 3} \left( {\Lambda \over
\Delta} \right)^{12} + \ldots }

We can now reconstruct this exact solution in perturbation theory by
simply evaluating the second derivative of $\cF_0$ at the critical
point up to a fixed number of loops. Given the perturbative expansion
\fpertt\ of $\cF_0$ in terms of a loop expansion of planar diagrams,
we should first compute $S=S_1=-S_2$ at the extremum. This gives a
series of the form
\eqn\SLAM{
{S\over \Delta^3}=
\left( {\Lambda \over \Delta} \right)^4
+ 6 \left( {\Lambda \over \Delta} \right)^8
+ 140 \left( {\Lambda \over \Delta} \right)^{12}
+ 4620 \left( {\Lambda \over \Delta} \right)^{16}
+ \ldots  }
Plugging this into $\partial^2\cF_0/\partial S^2$ gives us a
systematic approximation of the effective coupling $\tau$.  It is
instructive to see how the instanton expansion of $\tau$ computed from
the $n$-loop free energy of the matrix model for various $n$ gives a
sequence of series expansions gradually converging to the exact
result:
\eqn\taucomp{\eqalign{
\tau_{\rm 1-loop} & = 2 \log \left( {\Lambda \over \Delta} \right) \cr
\tau_{\rm 2-loop} & = 2 \log \left( {\Lambda \over \Delta} \right)
+ 20 \left( {\Lambda \over \Delta} \right)^4
+ 120 \left( {\Lambda \over \Delta} \right)^8
+ 1080 \left( {\Lambda \over \Delta} \right)^{12} + \ldots \cr
\tau_{\rm 3-loop} & = 2 \log \left( {\Lambda \over \Delta} \right)
+ 20 \left( {\Lambda \over \Delta} \right)^4
+ 538 \left( {\Lambda \over \Delta} \right)^8
+ 7816 \left( {\Lambda \over \Delta} \right)^{12} + \ldots \cr
\tau_{\rm 4-loop} & = 2 \log \left( {\Lambda \over \Delta} \right)
+ 20 \left( {\Lambda \over \Delta} \right)^4
+ 538 \left( {\Lambda \over \Delta} \right)^8
+ {62048 \over 3} \left( {\Lambda \over \Delta} \right)^{12} + \ldots \cr
& \vdots \cr
\tau_{\rm exact} & = 2 \log \left( {\Lambda \over \Delta} \right)
+ 20 \left( {\Lambda \over \Delta} \right)^4
+ 538 \left( {\Lambda \over \Delta} \right)^8
+ {62048 \over 3} \left( {\Lambda \over \Delta} \right)^{12}
%+ 922253 \left( {\Lambda \over \Delta} \right)^{16}
+ \ldots
}}

As an aside we point out that the condition $S_1=-S_2$, that naturally
emerges from minimizing the effective superpotential, means that from
the point of view of the matrix model we are dealing with a symmetric
filling of the two cuts. The exact solution to this model has
interesting properties and is further analyzed in Appendix A.
In particular there it is discussed that this model, as well as
its generalisation with asymmetric filling of the two cuts,
has a non-trivial scaling limit in the universality class
of the $c=1$ string.

Remembering the relation $S_i = g_s M_i$, we see that because
of the minus sign in \sss\ in the symmetric filling the number of
eigenvalues in the unstable cut (the maximum of the potential) is
negative.  This is clearly an unphysical solution and should be
interpreted as obtained by analytic continuation. In fact, if we put a
positive number of eigenvalues at an unstable critical point the
eigenvalue cut will not lie on the real axis but the cut will rotate
itself along the imaginary axis. (This can be seen by simply
analytically continuing $\F \to i\F$ in the Gaussian approximation.)
Instead of working with negative numbers it is perhaps better to think
of this solutions in terms of ``eigenvalue holes'' obtained by filling
the Dyson sea almost to the top of the potential.

Finally, let us point out that using matrix model results
we could also obtain other holomorphic quantities,
such as the SW periods ${\rm a}$ and ${\rm a}_D$.
For example, for the expectation values
$\langle \tr \Phi^k \rangle$ one finds a nice expression:
$$
\langle \tr \Phi^k \rangle = \oint x^k h
$$
written in terms of the 1-form $h=W''(x)dx/y$,
which can be interpreted as the smeared density of
the eigenvalues of the adjoint field $\Phi$ \refs{\DVthree,\CVlast}.
In particular, the case $k=1$ gives rise to the SW period ${\rm a}$.

%%%%%%%%%%%%%%%%%%%%%%%%%%%%%%%%%%%%%%%%%%%%%%%%%%%%%%%%%%%%%%%%%%%%%

\subsec{$\CN=1^*$ Theory}

As another illustration of the perturbative technique
in the matrix model applied to non-perturbative gauge theory,
we consider a massive deformation of $\CN=4$ gauge theory,
the so-called $\CN=1^*$ theory. In $\CN=1$ superspace
this theory is described by $U(N)$ gauge theory with
three adjoint chiral superfields and a tree-level superpotential:
\eqn\wtreeone{W_{{\rm tree}} = \tr \left( g \Phi_1 [ \Phi_2 , \Phi_3 ]
+ {m \over 2} \sum_{i=1}^3 \Phi_i^2 \right)}
For simplicity, we also assume that the eigenvalues of all
the Higgs fields are in the same classical vacuum (perturbation
theory around other vacua is discussed in Appendix B).
%Then, at the quantum level, the effective physics will be
%described by a single gluino bilinear (super)field $S$ and
%by the effective superpotential \weffa\
%%
%\eqn\weff{W_{\rm eff} (S) = \sum_i \Big[
%N_i S_i \log (S_i / \Lambda_i^3) - 2 \pi i \tau_0 S_i
%+ N_i {\partial F_{pert} (S) \over \partial S_i} \Big]}
%%
%Non-trivial part of this superpotential is the last term,
%which is expressed in terms of the genus zero free energy,
%$F_{pert} (S)$, of the corresponding matrix model.
Computing planar Feynman diagrams up to 3 loops in
the corresponding matrix model
we will be able to reproduce the leading terms in
the (exact) effective superpotential of $\CN=1^*$ theory.

{}From the topology of planar Feynman diagrams in this matrix model
it is easy to see that the free energy, $\CF_{0} (S)$,
has the following structure
%\foot{The same is true about any theory
%with interactions only up to degree three.}
%
\eqn\fpert{\CF_{0} (S)
= \sum_k c_{k+1} {g^{2k} \over m^{3k}} S^{k+2}}
Following the notations of \DVthree, henceforth we set $g=1$.
%(this can be always achieved by the appropriate rescaling of fields).

\ifig\perta{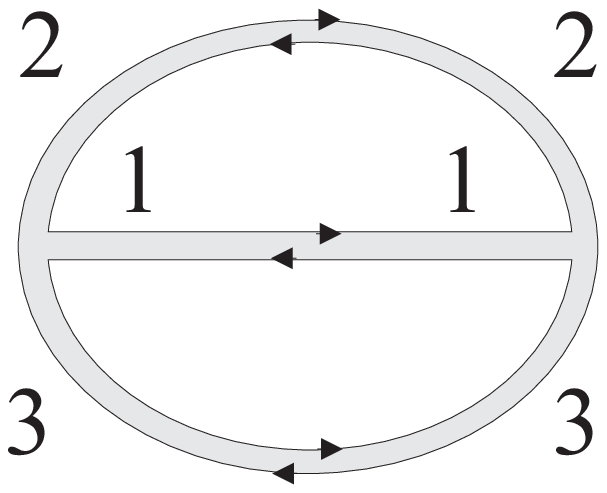}{30}{
Two-loop contribution to the free energy in the 3-matrix model
corresponding to the $\CN=1^*$ theory.  The numbers next to
propagators label the choice of one of the three matrix fields.}
%{\epsfxsize2.0in\epsfbox{perta.eps}}

Given the matrix model free energy $\CF_0 (S)$, one can compute the
effective superpotential $W_{{\rm eff}} (S)$ using the relation
\weffa.  Furthermore, integrating out the field $S$ in $W_{\rm eff}
(S)$ gives the effective superpotential as a function of the coupling
constants.  For the $\CN=1^*$ theory the answer can be computed
explicitly \DVthree\ by using the matrix model techniques developed in
\KKN.  Specifically, one obtains
\eqn\finwone{W_{\rm eff} = - {N m^3 \over 24} E_2 (\tau),}
where $\tau = \tau_0 / N$ and $E_2 (\tau)$ is the Eisenstein series.
This agrees with the analysis of \doet\ based on field theory dualities.
Up to an additive constant,
we can write the effective superpotential \finwone\ as a power
series in the variable $q = \exp (2 \pi i \tau)$:
%%
%\eqn\eisenstein{\eqalign{
%E_2 (\tau)
%& = 1 - 24 \sum_n {n q^n \over 1 - q^n} \approx \cr
%& \approx 1 - 24 q - 72 q^2 - 96 q^3 - 168 q^4 - 144 q^5 + \ldots
%}}
%%
%Therefore, up to an additive constant, the effective
%superpotential \finwone\ looks like
%
\eqn\wfinone{W_{\rm eff} = Nm^3 (q + 3 q^2 + 4 q^3 + 7q^4 + 6 q^5 +\ldots),}

Our goal is to reproduce this result by the perturbative
technique in the corresponding 3-matrix model
%with action \wtreeone.
%
\eqn\nstarmmodel{\int d \Phi\ \exp - \tr \Big(\Phi_1 [\Phi_2 , \Phi_3]
+ {m \over 2} \sum_{i=1}^3 \Phi_i^2 \Big).}
Namely, computing the planar Feynman diagrams up to three loops we
shall find numerical coefficients $c_k$ in the perturbative series
\fpert\ and, in particular, to check the first few coefficients in
eq. \wfinone.

The two-loop contribution to $\CF_{0}$ comes from the Feynman diagrams
of the type shown in \perta.  It is one of the diagrams that appears
in a simple 1-matrix model with cubic potential, see \twoloop.  The
second type of 2-loop diagrams in \twoloop\ does not appear here due
to the index structure of the cubic interaction.  Thus, we obtain the
two-loop coefficient $c_2=-1$.

\ifig\pertb{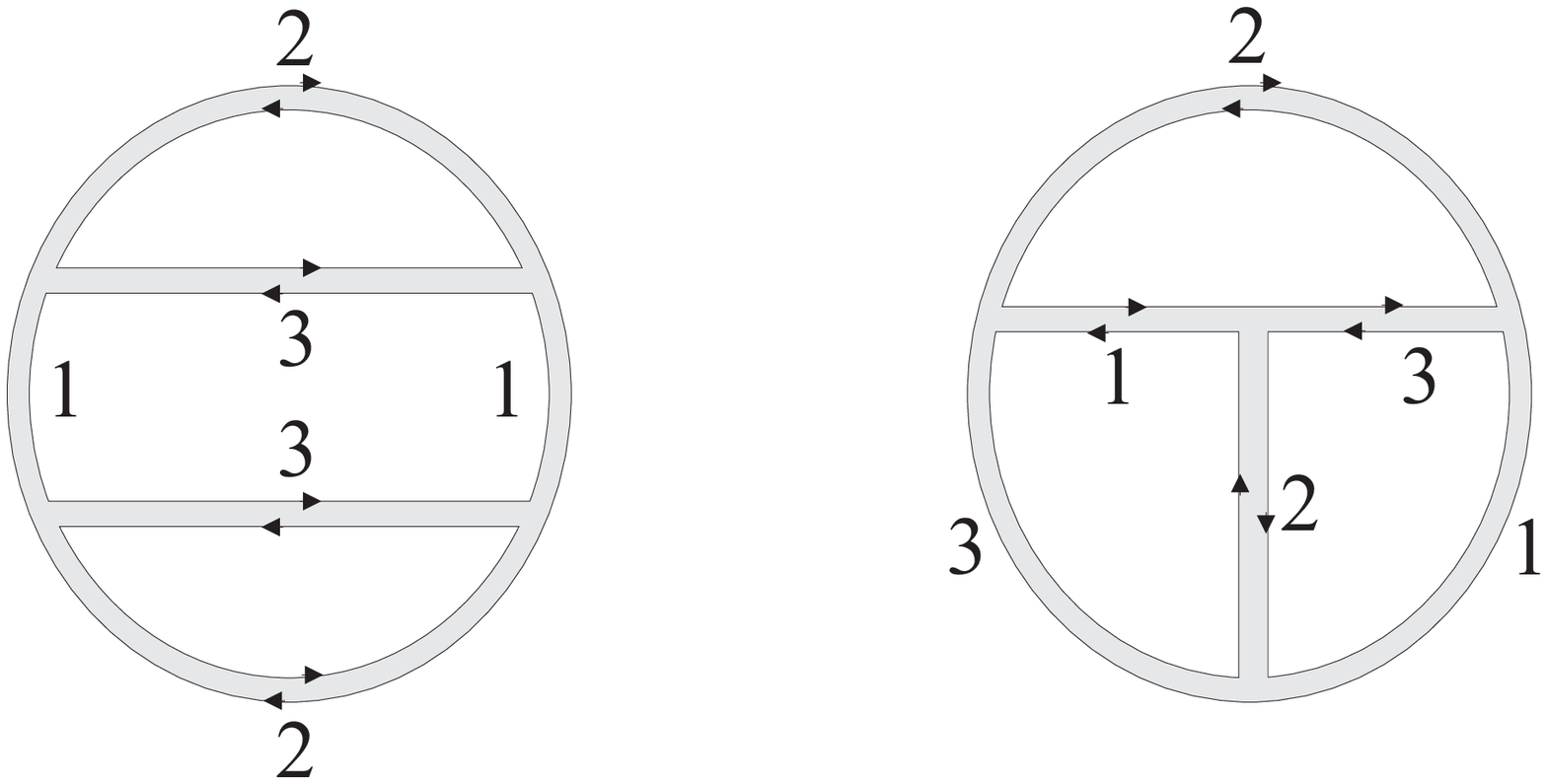}{80}{
Two types of 3-loop diagrams that contribute to $\CF_{0}$ with one of
the possible labeling of the propagators.}
%%{\epsfxsize4.0in\epsfbox{pertb.eps}}

At the next order, {\it i.e.} at three loops, there are two types of
diagrams which are presented in \pertb.  Taking into account also the
index structure of the diagrams one finds many different terms.
Adding all of them together gives $c_3 = 7/2$.

Summarising, up to three loops the perturbative expansion of $\CF_{0}
(S)$ has the form:
\eqn\fpertone{
\CF_{0} (S) = - {S^3 \over m^3}
+ {7 \over 2} {S^4 \over m^6} + \ldots }
Note, that the expansion we find indeed has the general structure
expected in \fpert. The relative minus signs in this expansion are due
to the interaction vertices with both positive and negative weight
arising from the commutator in \nstarmmodel.

Substituting \fpertone\ into \weffa\ we obtain the leading behavior of
the effective superpotential
\eqn\weffonea{
W_{\rm eff} = N S \log (S / m^3) - 2 \pi i \tau_0 S
- 3 N {S^2 \over m^3} + 14 N {S^3 \over m^6} + \ldots}
Now, integrating out the gluino field $S$
%means extremizing $W_{\rm eff} (S)$,
%then solving for $S(\tau)$ and substituting the result back into
%the effective superpotential. At the first step we find:
%%
%\eqn\wmin{
%W'_{\rm eff} (S) = N \log (S / \Lambda^3) + N - 2 \pi i \tau_0
%- 6 N {S \over m^3} + 42 N {S^2 \over m^6} + \ldots = 0}
%%
%Solving this quation for $S(\tau_0)$ we get
%(to the order $q^3$)
%$$
%S = \Lambda^3 q + 6 {\Lambda^6 \over m^3} q^2
%+ 12 {\Lambda^9 \over m^6} q^3 + \ldots
%$$
%%$$
%%S = \Lambda^3 \exp \big( {2 \pi i \tau_0 \over N} -1 \big)
%%+ 6 {\Lambda^6 \over m^3}
%%\exp 2 \big( {2 \pi i \tau_0 \over N} -1 \big)
%%+ \ldots
%%$$
%and plugging this back into \weffonea\
we obtain the final expression for the effective superpotential
\eqn\weffoneb{
W_{\rm eff}  = - N m^3 q - 3 N m^3 q^2 - 4 N m^3 q^3 + \ldots}
%
%\eqn\weffoneb{\eqalign{
%W_{\rm eff} = & N \Lambda^3 q (1 + 6 {\Lambda^3 \over m^3} q)
%\big[ \log q + \log (1 + 6 {\Lambda^3 \over m^3} q) \big] - \cr
%& - N (\log q + 1) \Lambda^3 q (1 + 6 {\Lambda^3 \over m^3} q)
%- 3 N {\Lambda^6 \over m^3} q + \ldots = \cr
%& = - N \Lambda^3 q - 3 N {\Lambda^6 \over m^3} q^2 + \ldots }}
%
%After fixing the scale $\Lambda=m$
The leading coefficients in this expression agree with the first
coefficients in the expansion of the exact answer \wfinone,
written in terms of the Eisenstein series $E_2 (\tau)$.
%(Of course, the overall sign --- more generally,
%a phase --- of $W_{\rm eff}$ is completely unphysical.)

Since we can do this calculation order by order, and since $n$-loop
diagrams give rise to $n$-instanton terms in $W_{\rm eff}$, it is
instructive to look at the higher order terms and to see how the
result depends on $n$.  For example, if we keep only the leading term in
the perturbative series $\CF_{0}$, the superpotential \weffonea\ looks
like:
\eqn\wefflead{
W_{\rm eff} = N S \log (S) - 2 \pi i \tau_0 S - 3 N S^2 }
This leads to
%the equation for $S$, {\it cf.} \wmin\
%$$
%\log (S/q) = 6 S
%$$
%which has the following solution
%$$
%S = q + 6 q^2 + 54 q^3 + 576 q^4 + 6750 q^5 + \ldots
%$$
%Note that we retain the terms of higher order in $q$,
%most of which can not be trusted in this approximation.
%Nevertheless, we proceed and substitute this result
%into \wefflead\ to find
the effective superpotential
\eqn\weffcrude{W_{\rm eff} =
Nm^3 (q + 3 q^2 + 18 q^3 + 144 q^4 + 1350 q^5 + \ldots)}
where we retained the terms of higher order in $q$,
most of which can not be trusted in this approximation.

If we compute perturbative free energy $\CF_{0}$ to three loops,
as we did above, we obtain the effective superpotential \weffoneb,
where one can trust three leading terms.
Moreover, the values of the higher order terms in \weffoneb\
are slightly ``improved'' compared to \weffcrude.
One can continue and do a similar calculation up to four loops
and so on. As a result, one finds a sequence of instanton expansions
which gradually approach the exact answer \wfinone:
\eqn\comparison{\eqalign{
W_{\rm 1-loop} & = Nm^3 {\underline q} \cr
W_{\rm 2-loop} & = Nm^3 \Big(
\underline{ q + 3 q^2} + 18 q^3 + 144 q^4 + 1350 q^5
+ {69984 \over 5} q^6 + {777924 \over 5} q^7 + \ldots \Big)
\cr
W_{\rm 3-loop} & = Nm^3 \Big(
\underline{q + 3 q^2 + 4 q^3} - 108 q^4 - 1548 q^5
- {43416 \over 5} q^6 + {345744 \over 5} q^7 + \ldots \Big)
\cr
W_{\rm 4-loop} & = Nm^3 \Big(
\underline{q + 3 q^2 + 4 q^3 + 7q^4} + 1212 q^5
+ {108384 \over 5} q^6 + {874744 \over 5} q^7 + \ldots \Big)
\cr
W_{\rm 5-loop} & = Nm^3 \Big(
\underline{q + 3 q^2 + 4 q^3 + 7q^4 + 6 q^5}
- {72516 \over 5} q^6 - {1657856 \over 5} q^7 + \ldots \Big)
\cr
W_{\rm 6-loop} & = Nm^3 \Big(
\underline{q + 3 q^2 + 4 q^3 + 7q^4 + 6 q^5 + 12 q^6}
+ 190976 q^7 + \ldots \Big)
\cr
& \vdots
\cr
W_{\rm exact} & = Nm^3 \Big(
\underline{q + 3 q^2 + 4 q^3 + 7q^4 + 6 q^5 + 12 q^6 + 8 q^7 + \ldots } \Big)
}}

Here, the underlined terms represent the exact terms in the instanton
expansion whose coefficients ``stabilize'' beyond a certain order.  It
is curious to note, that although all the numerical coefficients in
the exact superpotential $W_{\rm exact}$ are integer numbers, it is
not the case for the result obtained from a finite number of loops in
matrix perturbation theory.  Moreover, the $n$-loop approximation to
$W_{{\rm exact}}$ is not a modular form, and one can see from the
examples listed above that in the truncation to $n$ loops the mistake
in the $(n+1)$th coefficient is quite large. This emphasizes the fact
that the Montonen-Olive duality is not put in by hand in this
formalism, but rather is derived.  In this sense, we are going one
step beyond duality.

Note that we can express the $S$-duality of the $N=1^*$ theory
succinctly as the statement that the effective glueball superpotential
$W_{\rm eff}(S)$ is given by a {\it Legendre transform} of a modular
form, in this given by $E_2(\tau)$ (with $\tau=\tau_0/N$).

%%%%%%%%%%%%%%%%%%%%%%%%%%%%%%%%%%%%%%%%%%%%%%%%%%%%%%%%%%%%%%%%%%%%

\subsec{Massive Deformation of the Leigh-Strassler Model}

So far we considered only examples for which exact solution was
already known. This was helpful for establishing some confidence in
the perturbative technique since it did not rely on the existence of
the exact results, which we used only to verify the perturbative answer.
As we explained in the introduction, in most of the models
we don't have this luxury and, therefore, perturbative analysis
remains as the only tool for obtaining non-perturbative results,
such as instanton expansion of the effective superpotential.
Here, we consider one such model.

\ifig\perte{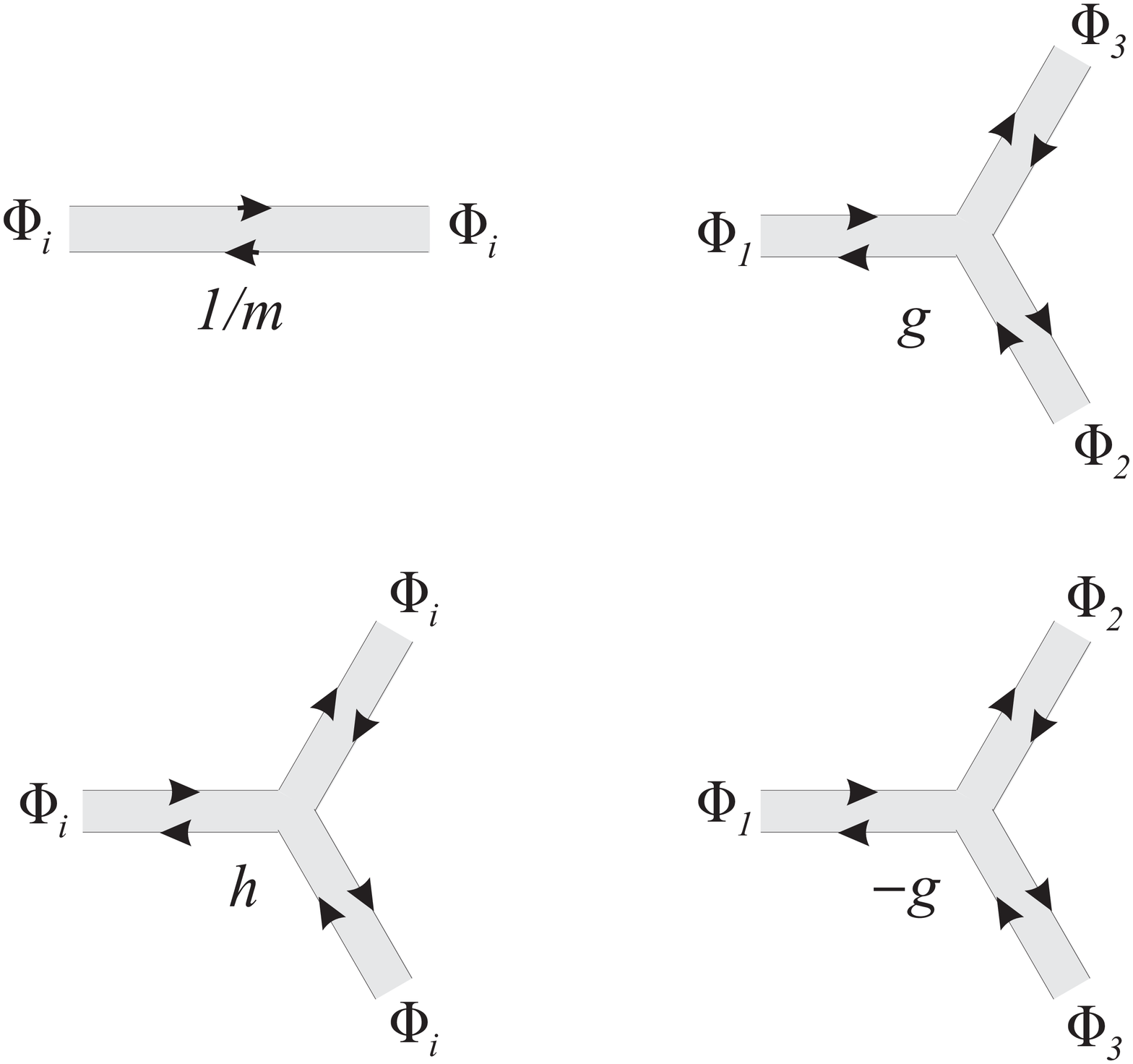}{70}{
The Feynman rules in the perturbative 3-matrix model corresponding to
the massive deformation of the Leigh-Strassler model.}
%{\epsfxsize4.0in\epsfbox{perte.eps}}

Specifically, we consider a Leigh-Strassler deformation \LS\
of the model discussed in the previous subsection:
\eqn\wone{
W_{\rm tree} = \tr \( g \Phi_1 [\Phi_2 , \Phi_3]
%\Phi_3 \Big( e^{i \pi \beta} \Phi_1 \Phi_2
%- e^{- i \pi \beta} \Phi_2 \Phi_1 \Big)
+ {h \over 3} \sum_{i=1}^3 \Phi_i^3
+ {m \over 2} \sum_{i=1}^3 \Phi_i^2 \) }
The corresponding 3-matrix model with action given by $W_{\rm
tree}(\Phi_i)$ can be solved in the large $M$ limit if either $g=0$ or
$h=0$, but the exact solution is not known when both deformation
parameters, $g$ and $h$, are non-zero.  On the other hand,
perturbation theory is very simple, with the Feynman rules summarized
in \perte.

At the 2-loop order, we find the following expression for the genus
zero free energy:
$$
\CF_{0} = {S^3 \over m^3} (2h^2 - g^2) + \ldots
$$
Substituting this into \weffa\ gives the effective superpotential
$$
W_{\rm eff} = N S \log (S / m^3) - 2 \pi i \tau_0 S
+ 3 (g^2 - 2h^2) N {S^2 \over m^3}  + \ldots
$$
Finally, extremizing it with respect to $S$
we obtain the value of the effective superpotential in the vacuum:
$$
W_{\rm eff} = Nm^3 \Big(q + 3 (g^2 - 2h^2) q^2 + \ldots \Big)
$$
%{\bf Is there any physical meaning of the curve $g^2 \sim 2h^2$ ?}
%Note, that the leading behavior of the effective superpotential
%does not depend on the deformation parameter $\beta$.
%This will not be true, however, after one takes into account
%higher order corrections.

The same technique applies to any $\CN=1$ theory
that admits a large $N$ limit. In particular,
one can systematically compute instanton corrections
to the effective superpotential in large class of
$\CN=1$ theories with any number of adjoint fields
and generic tree-level superpotentials.

%%%%%%%%%%%%%%%%%%%%%%%%%%%%%%%%%%%%%%%%%%%%%

\vskip 30pt

\centerline{\bf Acknowledgments}

We would like to thank M.~Aganagic, R.~Bousso, F.~Cachazo, S. J. Gates
Jr, M.~Mari\~no, A.~Marshakov, H.~Ooguri, S.~Theisen and K.~Zarembo
for valuable discussions.  The research of R.D. is partly supported by
FOM and the CMPA grant of the University of Amsterdam. S.G. is
supported by the Clay Mathematics Institute, RFBR grants 01-01-00549
and 02-01-06322.  V.A.K. is partly supported by European Union under
the RTN contracts HPRN-CT-2000-00122 and -00131.  C.V. is partly
supported by NSF grants PHY-9802709 and DMS-0074329.  We would like to
thank the Max Plank Institute in Potsdam (V.A.K.), Ecole Normale
Sup\'erieure (S.G.), and Harvard University (R.D.) for kind
hospitality during part of this work.

\vskip 30pt

%%%%%%%%%%%%%%%%%%%%%%%%%%%%%%%%%%%%%%%%%%%%%%%%%%%%%%%%%%%%%%%%%%%%%

\appendix{A}{Large $M$ Solution of the Two-Cut Matrix Model}

The results of perturbative expansions \CIV\ used in this section can
be reproduced, in accordance with the observations of \DVthree, from
the direct solution of the matrix model \PARTF, as was shown
in \DVone.  In this appendix we review, for the sake
of completeness, both the matrix model derivation as well
as the analytic form of the glueball superpotential
in terms of elliptic functions.
We take the cubic potential to be of the form:
\eqn\defS{
W(\Phi)=\tr \left( {1\over 4} \Phi-{1\over 3}\Phi^3 \right)
= \tr \left( \pm
\hf(\Phi\pm \hf)^2-{1\over 3}(\Phi\pm \hf)^3 \mp {1\over 12}\right) }
The last line is the expansion around each of the two symmetric
extrema of the potential. Note that we set here $\Delta=1$.

In terms of the eigenvalues, Using \EVL, we write the usual for the
one matrix model saddle point equation (SPE) in the large $M$ limit,
in terms of the eigenvalues,
\eqn\SPE{
x^2-{1\over 4} = 2 \l {-\!\!\!\!\!\!\int} du\rho(u) {1\over x-u},    }
where $\l=g_s M$ is the overall 't~Hooft coupling.
The two-cut solution can be found in terms of the analytical function
\eqn\ANAL{ \eqalign{
G(x)&=2\[\int_{x_1}^{x_2}+\int_{x_3}^{x_4}\] du\rho(u) {1\over
x-u}\cr &={1\over \l}\[ x^2-{1\over 4}-\sqrt{(x-x_1)(x-x_2)(x-x_3)(x-x_4)}\] } }
having the large $x$ asymptotics $ G(x\to\infty)=2{S_1+S_2\over \l x}$
and the
corresponding couplings on each of the two intervals $S_j =g_s M_j$, 
$j=1,2$, finite in the limit $g_s\to 0,$ $M, M_j \to \infty$.

The limits $x_i$, $i=1,2,3,4$ are defined by the large $x$ asymptotics:
\eqn\LIMITS{  \eqalign{   \sum_i x_i&=0  \cr
\sum_i x_i^2&=1  \cr
\sum_i x_i^3&= 12(S_1+S_2)   }}
and by the normalization condition for  the two intervals.
The latter is given in terms of the elliptic  integrals
\eqn\SOD{\eqalign{
S_{1}&={1\over 2\pi} \int_{x_1}^{x_2} dx\
\sqrt{(x_1-x)(x-x_2)(x-x_3)(x-x_4)} \cr
S_{2}&={1\over 2\pi}\int_{x_3}^{x_4} dx\
\sqrt{(x-x_1)(x-x_2)(x-x_3)(x-x_4)} }}
Let us now compute the free energy $\CF_{0}(S_1,S_2)={1\over M^2}\log Z$.
{}From the eigenvalue representation of the matrix model we
obtain the derivative of the free energy, amounting to the removal of
the eigenvalue at the edge of each cut:
$$
\p_{S_1}\CF_0 (S_1,S_2) = g_s^{-1}\( \CF_0 (S_1,S_2)-\CF_0 (S_1-g_s,S_2)\)=
 {1\over \l^2}W(x_1)+{2\over  M\l} \sum_{j\ne 1}\log(x_1-x_j),
$$
and a similar expression for $\p_{S_2}\CF_0 (S_1,S_2)$.
In terms of the eigenvalue density this gives:
$$
\eqalign{
& \l^2\p_{S_1}\CF_0 (S_1,S_2) = \cr
&  \quad
= W(x_1)+{1\over 2\pi i  }\[\oint_{x_1}^{x_2}
+\oint_{x_3}^{x_4}\]\sqrt{(x_1-x)(x-x_2)(x-x_3)(x-x_4)}\log(x_1-x)dx
}
$$
$$\eqalign{
&
\l^2\p_{S_2}\CF_0 (S_1,S_2) =\cr
& \quad = W(x_4)+{1\over 2\pi
i}\[\oint_{x_1}^{x_2}+\oint_{x_3}^{x_4}\]
\sqrt{(x_1-x)(x-x_2)(x-x_3)(x-x_4)}\log(x_4-x)dx}
$$
By expanding the contour of integration we pick up the contribution on
the logarithmic cut (apart from singularities at $x=\infty$
which we have to subtract in the matrix model framework). This gives:
\eqn\PERB{\eqalign{
\p_{S_1}\CF_0 (S_1,S_2)&=W(x_1)+  \Pi_1 +
{\rm subtractions}\ {\rm for}\ \L_0\to\infty
\cr
\p_{S_2}\CF_0 (S_1,S_2)&=W(x_4)+ \Pi_2 +
{\rm subtractions}\ {\rm for}\ \L_0\to\infty  }}
where $\L_0\to\infty$ is a cut-off and
\eqn\PERP{\eqalign{
 \Pi_1&={1\over\pi}\int_{-\L_0}^{x_1}
\sqrt{(x-x_1)(x-x_2)(x-x_3)(x-x_4)} dx \cr
\Pi_2& =-{1\over\pi}\int_{x_4}^{\L_0}
\sqrt{(x-x_1)(x-x_2)(x-x_3)(x-x_4)} dx  }}
are the dual periods. Formulas of this type appeared in \DAVID,
see also \DVone. In \CIV\ they follow from the analysis of the
Calabi-Yau geometry with flux. Using \SOD, \PERB\ and \PERP\ and
\LIMITS\ one finds the small $S_1,S_2$ expansion for the free energy
itself \fpertt\ from \CIV.

Let us note that the branch points are not necessarily placed on the
real axis. For a general complex $g_s$, they will choose their
positions according to the steepest decent in the eigenvalue
integral. For a real $g_s$ the stable cut will be on the real axis,
whereas the unstable cut will cross the real axis, having the complex
conjugated branch points. The situation when all branch points are on
the real axis corresponds to the analytical continuation in the
(originally positive) variables: $S_1>0,\ S_2<0$.

%%%%%%%%%%%%%%%%%%%%%%%%%%%%%%%%%%%%%%%%%%%%%%%%
\subsec{Symmetric Filling of Two Intervals }

Let us consider the case of the symmetric filling of two intervals
$x\in (b,a)$ and $x\in (-a,-b)$. It corresponds to the ``unphysical''
filling parameters $\hf g_s M=S=S_1=-S_2>0$, but nevertheless it will
reproduce the corresponding particular case of planar graph expansion
considered in the previous section. One can say that the two intervals
are filled by $M/2$ eigenvalues and $M/2$ ``holes'', respectively.  As
discussed in section 3.1, this case describes the $SU(2)$
Seiberg-Witten solution. We will also see yet another
way of obtaining $c=1$ noncritical string at a self-dual
radius from matrix models, when the endpoints of
the cuts approach each other.

The function $G(x)$, having the large $x$ asymptotics
$G(x\to\infty)=O(1/x^2)$, can be represented as
$$
\l\ G(x)=x^2-{1\over 4}-\sqrt{(x^2-a^2)(x^2-b^2)}
$$

The function $G(x)$, having the large $x$ asymptotics
$G(x\to\infty)=O(1/x^2)$, can be represented as
$$
\l\ G(x)=x^2-{1\over 4}-\sqrt{(x^2-a^2)(x^2-b^2)}
$$
The large $x$ asymptotics fixes one relation between $a$ and $b$:
$a^2+b^2=\hf$, and the normalization of the density $\int_b^a {dx\over
2\pi\l}\sqrt{(a^2-b^2)(x^2-b^2)}=1$ gives the relation (using \BF,
217.27\foot{beware of a mistake there: $g\to a$} and 361.01):
\eqn\SMA{\eqalign{   \l S&={1\over 2\pi} \int_b^a dx  
\sqrt{(a^2-x^2)(x^2-b^2)}\cr
&={a^3\over 6\pi}\left[(2-m) \BE - 2(1-m) \BK \right] }}
where where $\BK(m)$ and $\BE(m)$ are the elliptic integrals of the
I-st and II-nd kind, $a={1\over \sqrt{4-2m}}$ and the elliptic nome 
is $m=1-b^2/a^2$.  
The derivative of the free energy \PERB\ can be
calculated by the deformation of the contour to the dual period
correspoding to the interval $(-b,b)$ as the complete elliptic
integral 
\eqn\PERS{   \p_S\CF(S,-S)=
{2\over \l}\int_{-b}^b\sqrt{(x^2-a^2)(x^2-b^2)} dx }
However, the simplest quantity to calculate is actually the second
derivative of the free energy, which is to be identified with
the $\tau$-parameter of the SW curve. The latter can be seen
already in the form of \PERS. Indeed, by writing
$(x^2-a^2)(x^2-b^2)= (x^2-{1\over 4})^2-\L^4$,
where $\L^4={m^2\over 16(2-m)^2}$, we obtain 
\eqn\TAU{    \p_S^2\CF_0 (S,-S)={\p_m \p_S\CF(S,-S)\over \p_m S}
=4{\BK(1-m)/\BK(m)}\equiv 4\tau  }
We found the explicite elliptic parametrization of the free energy:
it is parametrized by $m$ which can be expressed through $S$
by means of \SMA.

Expanding \SMA\ and \TAU\ in powers of $\L^4={m^2\over 16(2-m)^2}$, we
get 
\eqn\SMAL{
S=\({\Lambda\over \sqrt{2}}\)^4 + 6 \({\Lambda\over \sqrt{2}}\)^8 +
140 \({\Lambda\over
\sqrt{2}}\)^{12} + 4620 \({\Lambda\over \sqrt{2}}\)^{16} + \ldots }
\eqn\TheisenTAU{
\tau = - {i \over \pi} \Big(
2 \log (\L^2/8) + {5 \over 2^3} \L^4
+ {269 \over 2^{10}} \L^8 + {1939 \over 3 \cdot 2^{12}}
\L^{12} + {922253 \over 2^{23}} \Lambda^{16}
+ \ldots \Big).}
The last is precisely the instanton expansion of the the SW coupling
constant, see {\it e.g.} \Klemm. It is not surprizing since the
numerator and denominator of \TAU\ coincide (up to the same factor)
with $\omega_D$ and $\omega$ from the formula (2.2) of \Klemm, whose
ratio defines $\tau$ of course.  Restoring the modulus $u$ and
rescaling $\L^2 \to 2 \L^2$, one can write this result in
conventions\foot{See also the footnote on page 3 in \Khoze.} of
ref. \Khoze, which also agree with our conventions used in section
3.1.  Specifically, one finds (up to an overall numerical factor):
\eqn\KhozeTAU{
\tau = \log (\L^2 / 4 u) + {5 \over 4} {\L^4 \over u^2}
+ {269 \over 2^{7}} {\L^8 \over u^4} + {1939 \over 3 \cdot 2^{7}}
{\L^{12} \over u^8} + {922253 \over 2^{16}} {\Lambda^{16} \over u^8} +
\ldots }
This is in agreement with \matrixtau,
as follows directly from the identification \udelrel\
of the $u$-variable with $\Delta^2/4$.

%To generate the small $S$ expansion of the free energy, let us expand
%\TAU\ and \SMA\ in power series in $\L={m^2\over 16(2-m)^2}$. 
%It is easy to check that the expansions coincide with the
%corresponding expansions \matrixtau\ and \SLAM\ from the section 3.1.
%%{\bf where $\Delta=4$, a discrepancy with the similar formula in
%%section 3.1: $\Delta=1$!}.
% 
It is not surprizing that inverting the series for $S$ plugging it
into \TAU\ and expanding in $S$ we obtain
\eqn\FREX{ \p_S^2\CF_0(S,-S) =2\log S +68 S +1500 S^2
+{142520\over 3}S^3+O(S^{4}) }
which coincides in the particular case $S=S_1=-S_2$ with the expansion
from \CIV\ quoted in section 3.1 (we put the dimensionful coupling
$\l=1$).

Using \SMA\ and \TAU\ we could also expand $\CF$ itself in terms of
the variable $q=e^{-\pi\tau}=e^{- \pi\BK(1-m)/\BK(m)}$, which will be
the instanton expansion for the corresponding $\CN=1$ SYM theory with
the $U(2)\to U(1)\times U(1)$ symmetry breaking cubic tree potential,
according to the recipe of \DVthree.

%%%%%%%%%%%%%%%%%%%%%%%%%%%%%%%%%%%%%%%%%%%%%%%%%%%%%%%%%%%%
\subsec{ $c=1$ critical regime }

In the context of the cubic potential matrix model,
there are two distinct ways of getting a $c=1$ non-critical string:
As noted in \ghov\ $c=1$ at self-dual radius
is equivalent to topological B-model on the deformed conifold, which
in turn has been shown to be equivalent to matrix model with quadratic
potential \DVone .  Thus in the theory we are dealing with, if
we zoom to the region near the critical points of the potential
we obtain a $c=1$ system at self-dual radius.  However, there
is another way of obtaining $c=1$ as well:  We can consider
the limit where the two ends of the cuts touch each other, which
again leads to a conifold geometry but now the vanishing cycle
is ``magnetic'' relative to the original ``electric'' cycle
of the matrix model.

Let us now look at this regime in more detail.  This
corresponds to $b\to 0$, when $m_1=1-m\to 0$ and
the two cuts in $F(x)$ merge into one. From  \SMA\ we obtain
in this limit
$$
\l S\simeq  {1\over 6\pi}-{1\over 8\pi}m_1\log(1/m_1) +O(m')
$$
which shows that this transition happens at $S_c={1\over 6\pi\l}$. The
free energy \TAU\ in this limit is
$$
 \p_S^2\CF_0(S,-S)\simeq {2\over\log {16/m_1}} +O(m_1)
$$
has, as a function of $\delta = \l(S_c-S)$, the typical behaviour of
the $c=1$ matter coupled to the 2D gravity \KaMi:
$$
\CF_0 (S)-\CF_0 (S_c)\simeq \l^2{\delta^2\over \log{1\over\delta} }
$$

%%%%%%%%%%%%%%%%%%%%%%%%%%%%%%%%%%%%%%%%%%%%%%%%%%%%%%%

\appendix{B}{Massive Vacua of $\CN=1^*$ Theory}

In this appendix we discuss matrix perturbation theory
for non-trivial massive vacua of $\CN=1^*$ theory,
corresponding to higher spin representations of $SU(2)$.
As we shall see, there are some novelties in perturbation theory,
which make these vacua conceptually similar to multi-cut matrix models.
In fact, the vacua we are going to discuss also correspond to
multi-cut matrix model \Dorey.
In both cases one finds (partial) gauge symmetry breaking
which leads to new fermionic ghost degrees of freedom.

In order to describe this more specifically, let us rewrite
the tree-level superpotential \wtreeone\ in $\CN=1^*$ theory 
in the following form:
\eqn\wtree{W_{\rm tree} = \tr \left( i [ \Phi_1, \Phi_2 ] \Phi_3
+ \sum_{i=1}^3 \Phi_i^2 \right) }
Supersymmetric vacua of the gauge theory correspond to
the critical points of this superpotential.
Thus, extremizing \wtree\ we find
\eqn\eom{[ \Phi_1, \Phi_2 ] = 2i \Phi_3}
plus two similar equations obtained by permutation of indices $1$, $2$, $3$.
One obvious solution corresponds to $\Phi_i =0$.
However, there are also some non-trivial solutions,
corresponding to $p$-dimensional representations of $SU(2)$.
In fact, suppose we start with a $U(N)$ gauge theory, with $N=pn$.
Then, we can take $n$ copies of such $p$-dimensional representations.
This leads to a partial breaking of gauge symmetry,
\eqn\gaugebreak{U(pn) \to U(n)}
Note that the rank of the gauge group has been reduced in this case
due to the fact that the irreducible representation we have taken
for vacuum configurations are not one dimensional.
The exact effective superpotential for all values
of $p$ is known \refs{\doet,\ADK,\Dorey},
and can be written in terms of the Eisenstein series $E_2 (\tau)$,
\eqn\weff{W_{\rm eff} = - {N p^2 \over 12} E_2 (\tau)}
very much like the superpotential in the for trivial vacuum, $p=1$.
The only novelty here is the relation between
$\tau$ and the bare coupling constant,
$$
\tau = p (p \tau_0 + k) / N
$$
In the effective field theory, this relation is set by
the tree-level term and the one-loop anomaly term in
the superpotential. The functional dependence on $\tau$,
on the other hand, is determined by matrix perturbative
expansion $\CF_{0}$ (around the corresponding vacuum).
Since for all values of $p$ we have the same functional
dependece on $\tau$ --- given by the Eisenstein series --- we
conclude that $\CF_{0}$ should be the same for all vacua,
{\it i.e.} for all values of $p$:
\eqn\fpertb{\CF_{0} = - S^3 + {7 \over 2} S^4 + \ldots}

In order to reproduce this result directly by perturbative techniques
in matrix model, we have to expand the superpotential \wtree\
near a vacuum:
$$
\Phi_1 \to X + \Phi_1, \quad
\Phi_1 \to Y + \Phi_2, \quad
\Phi_1 \to Z + \Phi_3
$$
where $X$, $Y$, and $Z$ solve \eom:
\eqn\xyzsutwo{[X,Y] = 2i Z, \quad {\rm etc.}}
Substituting this into \wtree\ we find:
\eqn\wtreexyz{W_{\rm tree} = \tr \left( i [ \Phi_1, \Phi_2 ] \Phi_3
+ \sum_{i=1}^3 \Phi_i^2
+ i X [\Phi_2, \Phi_3] + i Y [\Phi_3, \Phi_1] + i Z [\Phi_1, \Phi_2]
\right)}

Let us consider a specific case, corresponding to $p=2$.
In this case, we have the following gauge symmetry
breaking pattern:
\eqn\ptwocase{ U(2M) \to U(M) }
Hence, it is convenient to write all the matrix variables
in terms of $M \times M$ blocks.
Specifically, we take (it is easy to check that this
is indeed a solution to \eom):
$$
X = \pmatrix{ 0 & {\bf 1} \cr {\bf 1} & 0}, \quad
Y = i \pmatrix{ 0 & - {\bf 1} \cr {\bf 1} & 0}, \quad
Z = \pmatrix{ {\bf 1} & 0 \cr 0 & - {\bf 1}}
$$
and for each hermitian matrix $\Phi_i$ we introduce the notation
\eqn\blockform{
\Phi = {1 \over 2}
\pmatrix{A^+ + A^- & D + i F \cr
D - i F & A^+ - A^- }}
where $A^{\pm}$, $D$, and $F$ are $M \times M$ matrices.
Using this decomposition for all of the three matrix fields $\Phi_i$,
we get in total $3 \times 4 = 12$ matrices of size $M \times M$:
\eqn\fieldsa{
A_1^{\pm}, \quad D_1, \quad F_1, \quad
A_2^{\pm}, \quad D_2, \quad F_2, \quad
A_3^{\pm}, \quad D_3, \quad F_3}
However, the gauge symmetry breaking \ptwocase\ suggests
that $3M^2$ degrees of freedom can be gauge fixed to zero,
so that effectively we should end up only with 9 matrix fields.
This is precisely what one finds.

Rewriting \wtreexyz\ in terms of $M \times M$ matrices
gives the following quadratic (mass) terms
\eqn\wquadfirst{\eqalign{
W_{\rm quadr} & = \tr \Big( {1 \over 2} \sum_i \left( A_i^+\right)^2
+ {1 \over 2} D_1^2 + {1 \over 2} F_2^2
+ {1 \over 2} \left( A_3^- \right)^2
+ D_1 F_2 + F_2 A_3^- - D_1 A_3^- + \cr
& + {1 \over 2} (D_2 - F_1)^2
+ {1 \over 2} (A_2^- - F_3)^2
+ {1 \over 2} (A_1^- + D_3)^2 \Big)
}}
Here, the fields in the first line have non-degenerate mass matrix.
However, the fields in the second line appear only in certain
linear combinations. Hence, their orthogonal combinations,
\eqn\zeromodes{\eqalign{
& D_2 + F_1 \cr
& A_2^- + F_3 \cr
& A_1^- - D_3
}}
represent massless directions and can be potentially dangerous
in the matrix integral.
In fact, these are simply the usual Goldstone zero-modes
which can be removed by gauge fixing.
We choose the following gauge, suggested by \zeromodes:
\eqn\gauge{\eqalign{
& D_2 = - F_1  \cr
& A_2^- = - F_3  \cr
& A_1^- = D_3
}}
This eliminates three out of twelve $M \times M$ matrices.
For example, if we choose to eliminate $D_2$, $A_2^-$ and $A_1^-$,
we end up with nine bosonic matrices:
\eqn\fieldsb{
A_1^{+}, \quad D_1, \quad F_1, \quad
A_2^{+}, \quad F_2, \quad
A_3^{+}, \quad A_3^{-}, \quad D_3, \quad F_3}

Next, we should introduce fermionic ghost fieldss $B$, $C$.
In order to do this, we note that under $SU(2M)$ gauge
transformation the matrix fields $\Phi_i$ transform as:
$$
\delta \Phi \sim [\Phi , C]
$$
Again, we write $C$ in the $2 \times 2$ block form,
similar to \blockform:
\eqn\ghostblock{
C = {1 \over 2}
\pmatrix{C_A & C_D + i C_F \cr
C_D - i C_F & - C_A}}
Applying the gauge transformation to \gauge\
and using the standard Faddeev-Popov method,
one finds the action for the ghost fields
$B_{\alpha} , C_{\alpha}$, where we introduced
a new index notation $\alpha = A,D,F$.
Straighforward, but slightly technical calculation gives:
\eqn\ghostw{\eqalign{ W_{\rm ghost} & =
\tr \Big( 8 i B_A C_A - 4i B_D C_D - 4i B_F C_F + \cr
& + {1 \over 2} B_A \Big[
2i C_A (D_1 - F_2) + 2i (D_1 - F_2) C_A +
+ C_D ( - A_2^+ - i D_3) + C_F (- A_1^+ - i F_3) \cr
& + (A_2^+ - i D_3) C_D + (A_2^+ - i F_3) C_F \Big] + \cr
& + {1 \over 2} B_D \Big[
2 C_A ( i D_3  - A_2^+) + 2 (A_2^+ + i D_3) C_A
+ C_D ( i F_2 - i A_3^-) + C_F (- A_3^+ + i F_1) + \cr
& + (i F_2 - i A_3^-) C_D + (i F_1 + A_3^+) C_F \Big] + \cr
& + {1 \over 2} B_F \Big[
2 C_A ( i F_3  - A_1^+) + 2 (A_1^+ + i F_3) C_A
+ C_D ( i F_1 + A_3^+) + C_F (- iD_1 - i A_3^-) + \cr
& + (i F_1 - A_3^+) C_D + (-iD_1 - i A_3^-) C_F \Big] \Big)
}}

Summarising, in the case of $p=2$ we find a $(9+6)$-matrix
model, that is a matrix model with 9 bosonic and 6 fermionic
(ghost) fields,
\eqn\fieldsc{\eqalign{
{\rm Bosonic~}~ \quad & : \quad
A_1^{+}, \quad D_1, \quad F_1, \quad
A_2^{+}, \quad F_2, \quad
A_3^{+}, \quad A_3^{-}, \quad D_3, \quad F_3 \cr
{\rm Fermionic~}~ \quad & : \quad
B_A, \quad B_D, \quad B_F, \quad
C_A, \quad C_D, \quad C_F
}}
and with the following action
\eqn\wtot{W_{\rm tree} = W_{\rm quadr} + W_{\rm cubic} + W_{\rm ghost}}
where the ghost action is given by \ghostw.
The quadratic terms of the bosonic action are
given by \wquadfirst:
$$
\eqalign{
W_{\rm quadr} = & \tr \Big( {1 \over 2} \sum_i \left( A_i^+\right)^2
+ 2 F_1^2 + 2 F_3^2 + 2 D_3^2 + \cr
& + {1 \over 2} D_1^2 + {1 \over 2} F_2^2
+ {1 \over 2} \left( A_3^- \right)^2
+ D_1 F_2 + F_2 A_3^- - D_1 A_3^- \Big)
}
$$
while the cubic interactions read
\eqn\wcubic{\eqalign{ W_{\rm cubic} & = \tr \Big(
{i \over 4} \Big(
[A_1^+ , A_2^+ ] A_3^+
+ \left( [F_1 , F_2] - [D_1, F_1] - [D_3 , F_3] \right) A_3^+ + \cr
& + \( [D_3 , D_1] + [F_3, F_1] \) A_2^+
+ \left( [D_3 , F_1] + [F_2, F_3] \right) A_1^+
- [A_1^+ , F_3 ] A_3^- + \cr
& + [D_3 , A_2^+ ] A_3^-
+ i \( -2 F_1^2  - D_1F_2 -F_2D_1 \) A_3^- + \cr
& + i \left( -2 F_3^2 D_1 + F_1 F_3 D_3 + F_1 D_3 F_3
+ 2 F_2 D_3^2  + F_1 F_3 D_3 + F_1 D_3 F_3 \right)
\Big) \Big)
}}

Computation of the planar Feynman diagrams in this matrix model
is expected to reproduce the perturbative expansion of the free
energy \fpertb. We will not pursue it further in this paper.

\listrefs
\end